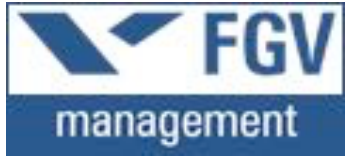

# FUNDAÇÃO GETÚLIO VARGAS

## CURSO PROGRAMA FGV MANAGEMENT MBA EM GERÊNCIA DE PROJETOS

# MODELO DE MATURIDADE EM GERENCIAMENTO DE RISCOS EM PROJETOS

**DANIEL MARSSIÉRE BIRCHAL**

**JOÃO MÁRCIO ABIJAODI**

**PAULO HENRIQUE ABREU**

**RICARDO MAGNO SANTOS ANTUNES**

**ROGÉRIO DO CARMO PEIXOTO**

**Dezembro/ 2007**

**DANIEL MARSSIÉRE BIRCHAL**

**JOÃO MÁRCIO ABIJAODI**

**PAULO HENRIQUE ABREU**

**RICARDO MAGNO SANTOS ANTUNES**

**ROGÉRIO DO CARMO PEIXOTO**

# MODELO DE MATURIDADE EM GERENCIAMENTO DE RISCOS EM PROJETOS

Trabalho apresentado ao curso MBA em Gerência de Projetos, Pós-Graduação *lato sensu*, da Fundação Getulio Vargas, como requisito parcial para a obtenção do grau de Especialista em Gerência de Projetos.

**Orientador: Prof. Carlos Salles**

**Belo Horizonte**
**Dezembro/ 2007**

# FOLHA DE APROVAÇÃO

O Trabalho de Conclusão de Curso da Fundação Getúlio Vargas - Programa FGV Management MBA em Gerência de Projetos, elaborado por Daniel Marssiére Birchal; João Márcio Abijaodi; Paulo Henrique Abreu; Ricardo Magno Santos Antunes e Rogério do Carmo Peixoto, sobre o título: *"Modelo de maturidade em gerenciamento de riscos em projetos"* e aprovado pela Coordenação Acadêmica do curso de MBA em Gerência de Projetos, foi aceito como requisito parcial para a obtenção do certificado do curso de pós-graduação, nível de especialização do Programa FGV Management.

Belo Horizonte, 10 de Dezembro de 2007

_______________________________________________

**Carlos A. C. Salles Jr.**
Coordenador Acadêmico Executivo e Prof. Orientador

**TERMO DE COMPROMISSO**

O(s) aluno(s): Daniel Marssiére Birchal; João Márcio Abijaodi; Paulo Henrique Abreu; Ricardo Magno Santos Antunes e Rogério do Carmo Peixoto, abaixo assinado(s), do curso de MBA em Gerência de Projetos, Turma GP10BH do Programa FGV Management, realizado nas dependências da BI Minas, no período de 10/03/2006 a 17/06/2007, declaram que o conteúdo do Trabalho de Conclusão de Curso, intitulado: *Modelo de Maturidade em Gerenciamento de Riscos em Projetos*, é autêntico, original e de sua autoria exclusiva.

Belo Horizonte, 10 de Dezembro de 2007

_______________________________
Daniel Marssiére Birchal

_______________________________
João Márcio Abijaodi

_______________________________
Paulo Henrique Abreu

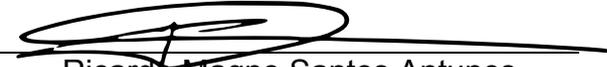
_______________________________
Ricardo Magno Santos Antunes

_______________________________
Rogério do Carmo Peixoto

**DEDICATÓRIA**

Este trabalho é dedicado aos mestres, que nos deram o
devido embasamento teórico.

**AGRADECIMENTOS**

Aos queridos familiares e amigos, que souberam entender
(e apoiar) nossa ausência, no decorrer do curso.

# RESUMO

A globalização alimentada pela explosão tecnológica iniciada no fim do século passado propulsionou o mundo a mudar cada dia mais rápido. A única certeza de hoje é a incerteza do amanhã. *Risco* é definido como incerteza onde uma, ou várias, causa(s) composta por probabilidade de ocorrência pode gerar determinado impacto ou conseqüência (ameaça se for negativo e oportunidade se for positivo para determinado objetivo). A gerência de riscos constitui-se da cultura, processos e procedimentos de uma organização ou indivíduo no trato das incertezas visando à minimização de ameaças e maximização das oportunidades quanto ao alcance de um objetivo estabelecido. O *"Modelo de maturidade em gerenciamento de riscos em projetos"*, proposto neste trabalho, objetiva mensurar a capacidade e habilidade de organizações gerenciarem os riscos envolvidos em projetos, dentro de uma metodologia de gêrencia de riscos praticada.

**Palavras-chave:** projetos; riscos; gerenciamento de riscos em projetos; modelo de maturidade.

# ABSTRACT

The globalization feeded by the technology explosion that begans in the end of the last century, started the world to change faster every day. The only today's certain is the tomorrow's uncertain. Risk is defined as uncertain where one or many causes composed of ocurrence probality can generate an impact or consequence (threat if negative and oportunity if positive, to a determinated goal). The Risk Management is composed of culture, procedure and process of an organization or individual care of uncertain, aiming to minimize threats e maximizing the oportunities, to reach a desired goal. The *"Risk maturity model in projects"* proposed on this document, wants to measure the organizations capacity and skills to manage the riks involved in projects when adopting a generic risk management methodology.

**Key-words:** projects; risks; project risk management; maturity model.

# LISTA DE FIGURAS

# LISTA DE QUADROS

# LISTA DE APÊNDICES

# LISTA DE ANEXOS

# LISTA DE ABREVIATURAS E SIGLAS

**ABGP** - Associação Brasileira em Gerenciamento de Projetos

**CMM** - Capability Maturity Model

**CMMI** - Capability Maturity Model Integration

**EAP** - Estrutura Analítica do Projeto

**ICB** - IPMA Competence Baseline

**IEC** - International Electrotechnical Commission

**IMPA** - International Management Project Association

**IPMA** - Internacional Project Management Association
Associação Internacional de Gerenciamento de Projetos

**IPPD** - Integrated Product and Process Development

**ISO** - International Organization for Standardization

**MMGP** - Modelo de Maturidade em Gerenciamento de Projetos

**NBR** - Normas Brasileiras

**NCB** - National Competence Baseline

**OGC** - Office of Government Commerce

**P2MM** - Prince2 Maturity Model

**P3M3** - Portfolio, Programme & Project Management Maturity Model

**PDCA** - Plan, Do, Check, Act

**PMBoK** - Project Management Body of Knowledge
Corpo de Conhecimento em Gerenciamento de Projetos

**PMI** - Project Management Institute

**PMMM** - Project Management Maturity Model

**PMO** - Project Management Office
Escritório de Projetos

**PRINCE2** - Projects in Controlled Environments

**RBC** - Referencial Brasileiro de Competências em Gerenciamento de Projetos

**SCAMPI** - Standard CMMI Appraisal Method for Process Improvement

**SE** - Software Engineering

**SEI** - Software Engineering Institute

**SPICE** - Software Process Improvement and Capability dEtermination

**SS** - Supplier Sourcing

# SUMÁRIO

# 1 INTRODUÇÃO

## 1.1 Considerações iniciais

A globalização alimentada pela explosão tecnológica iniciada no fim do século passado propulsionou o mundo a mudar cada dia mais rápido. Paradigmas caem por terra a todo o momento, verdades absolutas tornam-se mentiras enquanto novas informações surgem a todo o momento de todas as partes. A única certeza de hoje é a incerteza do amanhã. Neste ambiente instável e mutável estão milhões de organizações em todo mundo, em que trabalham bilhões de pessoas, onde uma visão correta do futuro fará toda a diferença entre crescer ou falir. A chave para a sobrevivência ou crescimento destas organizações, sem dúvida, esta na habilidade em lidar com as incertezas no mundo atual, em prever o futuro.

Defini-se risco como incerteza onde uma, ou várias, causa(s) composta por probabilidade de ocorrência pode gerar determinado impacto ou conseqüência. O risco pode ser classificado como ameaça se seu impacto for negativo e como oportunidade se este impacto for positivo a determinado objetivo.

A gerência de riscos constitui-se da cultura, processos e procedimentos de uma organização ou indivíduo no trato das incertezas visando à minimização de ameaças e maximização das oportunidades quanto ao alcance de um objetivo estabelecido.

O modelo de maturidade em gerencia de risco proposto neste trabalho objetiva mensurar a capacidade e habilidade de organizações gerenciarem os riscos envolvidos em seus negócios e ou projetos dentro de uma metodologia de gerência de riscos praticada.

A partir do estudo dos principais modelos de maturidade para gerenciamento de projetos, programas e *portfólios* e de desenvolvimento de software utilizados atualmente estabeleceu-se o modelo de maturidade específico para a metodologia de gerenciamento de riscos apresentado neste trabalho. Esta, fruto da compilação das melhores práticas de diferentes metodologias praticadas pelo mercado a fim de abranger todos os níveis de maturidade.

# 2 FUNDAMENTAÇÃO TEÓRICA

## 2.1 Melhores práticas em projetos e organizações mundiais voltadas ao gerenciamento de projetos (PMI e IPMA)

Atualmente, existem várias organizações e associações em todo o mundo, focadas em gerenciamento de projetos, e estes, dos mais variados contextos. Entre elas, destacamos:

- o PMI (Project Management Institute);
- o IPMA (International Management Association).

### 2.1.1 PMI (Project Management Institute)

O PMI tem maior concentração de associados nos continentes Americanos. Sua maior referência bibliográfica é o PMBoK (Project Management Body of Knowledge ou Corpo de Conhecimento em Gerenciamento de Projetos), que fornece um direcionamento de melhores práticas para a gestão de projetos.

### 2.1.2 As nove áreas de conhecimento

O PMBok abrange nove áreas de conhecimento: Integração: Escopo, Tempo, Custo, Qualidade, Recursos Humanos, Comunicação, Risco e Aquisições.

### 2.1.2.1 Escopo

Descreve os processos envolvidos na verificação de que o projeto inclui todo o trabalho necessário, e apenas o trabalho necessário, para que seja concluído com sucesso. Ele consiste nos processos de gerenciamento de projetos:

- Planejamento do escopo;

- Definição do escopo;

- Criar a Estrutura Analítica do Projeto;

- Verificação do escopo e

- Controle do escopo.

**2.1.2.2 Tempo**

Descreve os processos relativos ao término do projeto no prazo correto. Ele consiste nos processos de gerenciamento de projetos:

- Definição da atividade;

- Eqüenciamento de atividades;

- Estimativa de recursos da atividade;

- Estimativa de duração da atividade;

- Desenvolvimento do cronograma e

- Controle do cronograma.

**2.1.2.3 Custo**

Descreve os processos envolvidos em planejamento, estimativa, orçamentação e controle de custos, de modo que o projeto termine dentro do orçamento aprovado. Ele consiste nos processos de gerenciamento de projetos:

- Estimativa de custos;

- Orçamentação e

- Controle de custos.

### 2.1.2.4 Qualidade

Descreve os processos envolvidos na garantia de que o projeto irá satisfazer os objetivos para os quais foi realizado. Ele consiste nos processos de gerenciamento de projetos:

- Planejamento da qualidade;

- Realizar a garantia da qualidade e

- Realizar o controle da qualidade.

### 2.1.2.5    Recursos humanos

Descreve os processos que organizam e gerenciam a equipe do projeto. Ele consiste nos processos de gerenciamento de projetos:

- Planejamento de recursos humanos;

- Contratar ou mobilizar a equipe do projeto;

- Desenvolver a equipe do projeto e

- Gerenciar a equipe do projeto.

### 2.1.2.6 Comunicação

Descrevem os processos relativos à geração, coleta, disseminação, armazenamento e destinação final das informações do projeto de forma oportuna e adequada. Ele consiste nos processos de gerenciamento de projetos:

- Planejamento das comunicações;

- Distribuição das informações;

- Relatório de desempenho e

- Gerenciar as partes interessadas.

### 2.1.2.7 Risco

Engloba os processos necessários para garantir a correta identificação, análise, e resposta aos riscos do projeto; maximizando os efeitos positivos e minimizando a conseqüência de efeitos negativos. Inclui:

- Gerenciamento do plano do risco;

- Identificação do risco;

- Análise qualitativa do risco;

- Análise quantitativa do risco;

- Planejamento da resposta ao risco;

- Monitoração e controle do risco;

### 2.1.2.8 Integração

Descreve os processos e as atividades que integram os diversos elementos do gerenciamento de projetos, que são identificados, definidos, combinados, unificados e coordenados dentro dos grupos de processos de gerenciamento de projetos. Ele consiste nos processos de gerenciamento de projetos:

- Desenvolver o termo de abertura do projeto;

- Desenvolver a declaração do escopo preliminar do projeto;

- Desenvolver o plano de gerenciamento do projeto;

- Orientar e gerenciar a execução do projeto;

- Monitorar e controlar o trabalho do projeto;

- Controle integrado de mudanças e

- Encerrar o projeto.

### 2.1.2.9 Aquisições

Descreve os processos relativos à realização do gerenciamento de riscos em um projeto. Ele consiste nos processos de gerenciamento de projetos:

- Planejamento do gerenciamento de riscos;

- Identificação de riscos;

- Análise qualitativa de riscos;

- Análise quantitativa de riscos;

- Planejamento de respostas a riscos e

- Monitoramento e controle de riscos.

### 2.1.3 Projetos, programas e portfólios

As referências do PMBok, podem ser utilizadas para o Gerenciamento tanto de Projetos, como Programas e Portfólios.

### 2.1.3.1 O que é projeto

> *Processo único, consistindo de um grupo de atividades coordenadas e controladas com datas para início e término, empreendido para alcance de um objetivo conforme requisitos específicos, incluindo limitações de tempo, custo e recursos.* (NBR, 10006)

*"Um empreendimento temporário, com objetivo de criar um produto, serviço ou resultado único."* (PMBok, 2004, p. 5)

Projetos tornaram-se um importante instrumento de mudança e desenvolvimento nas organizações. As principais mudanças organizacionais e as iniciativas para gerar vantagens competitivas têm sido executadas, em sua maior parte, através de projetos organizacionais. Dessa forma, a disciplina gerenciamento de projetos vem ganhando destaque dentro dos modelos de administração e tem-se transformado num fator relevante para prover velocidade, robustez, consistência e excelência operacional na consecução de projetos.

### 2.1.3.2 O que é programa

*"Um programa é um grupo de projetos gerenciados de forma coordenada, visando obter benefícios difíceis de serem obtidos quando gerenciados isoladamente"* (PMBok, 2004, p.16).

Programas podem incluir elementos de trabalho relacionado fora do escopo dos projetos distintos no programa. Por exemplo: o programa de um novo modelo de carro pode ser subdividido em projetos para o *design* e as atualizações de cada componente principal (por exemplo: transmissão, motor, interior, exterior) enquanto a fabricação continua na linha de montagem.

Muitas empresas de produtos eletrônicos possuem gerentes de programas responsáveis tanto pelos lançamentos (projetos) de produtos específicos quanto pela coordenação de vários lançamentos durante um período de tempo (uma operação contínua).

Os programas também envolvem uma série de empreendimentos repetitivos ou cíclicos. Por exemplo:

- As empresas de serviços públicos freqüentemente falam de um "programa de obras" anual, uma série de projetos desenvolvidos com base em esforços anteriores.

- Muitas organizações sem fins lucrativos possuem um "programa de arrecadação de fundos" para obter apoio financeiro envolvendo uma série

de projetos distintos, como uma campanha para atrair novos sócios ou um leilão.

- A publicação de um jornal ou uma revista também é um programa em que cada problema específico é gerenciado como um projeto.

Este é um exemplo de casos que operações genéricas que podem se tornar um "gerenciamento por projetos" (SEÇÃO 1.3).

Ao contrário do gerenciamento de projetos, o gerenciamento de programas é o gerenciamento centralizado e coordenado de um grupo de projetos para atingir os objetivos e benefícios estratégicos do programa.

### 2.1.3.3 O que é *portfólio*

Um *portfólio* é um conjunto de projetos ou programas e outros trabalhos agrupados para facilitar o gerenciamento eficaz desse trabalho, a fim de atender aos objetivos de negócios estratégicos. Os projetos ou programas no *portfólio* podem não ser necessariamente interdependentes ou diretamente relacionados. É possível atribuir recursos financeiros e suporte com base em categorias de risco/premiação, linhas de negócios específicas ou tipos de projetos genéricos, como infra-estrutura e melhoria dos processos internos.

As organizações gerenciam seus *portfólios* com base em metas específicas. Uma meta do gerenciamento de *portfólios* é maximizar o valor do *portfólio* através do exame cuidadoso dos projetos e programas candidatos para inclusão no *portfólio* e da exclusão oportuna de projetos que não atendam aos objetivos estratégicos do *portfólio*. Outras metas são equilibrar o *portfólio* entre investimentos incrementais e radicais e para o uso eficiente dos recursos. Os diretores e equipes de gerenciamento da diretoria normalmente assumem a responsabilidade de gerenciar os *portfólios* para uma organização.

### 2.1.4 O que é gerência de projetos?

*"Gerência de projetos é a aplicação de conhecimentos, habilidades, ferramentas e técnicas nas atividades do projeto com o objetivo de atender os requisitos do projeto"* (PMBok, 2004, p.9 (PMI)).

### 2.1.4.1 Escritório de projetos - PMO

No tocante à estrutura organizacional voltada especificamente ao gerenciamento de projetos é também muito importante destacar o papel vital do PMO (*Project Management Office* – Escritório de Projetos). O PMO pode ser definido com a estrutura organizacional estabelecida para apoiar os gerentes e as equipes de projetos na implementação de princípios, práticas, metodologias, ferramentas e técnicas para o gerenciamento de projetos (DAI & WELLS, 2004). O PMO pode apoiar de forma significativa e contundente a transformação das estratégias da organização em projetos e planos de ação através de um adequado e eficiente gerenciamento de projetos.

Pesquisas, segundo Roolins (2003), apontam para a existência de mais de 50.000 Escritórios de Projetos existentes nos Estados Unidos. Verzuh (1999) argumenta que se uma organização desenvolve projetos de forma esporádica, não há a necessidade de desenvolver de forma sistemática habilidades para as iniciativas de projetos.

Todavia, se uma organização dedica grande parte de sua energia à implementação de projetos, uma abordagem não-estruturada e disciplinada para o gerenciamento de projetos conduz a ineficiências que podem ser danosas às organizações. Com um grande número de projetos sendo gerenciada, a necessidade da presença de um PMO torna-se evidente.

### 2.1.5 IPMA (International Project Management Association)

A Associação Internacional de Gerenciamento de Projetos (IPMA - Internacional Project Management Association) foi criada em Viena, Suíça, em 1965. Atualmente é mais aceita e conhecida em países da Europa, mas começa a crescer rapidamente no continente americano.

Cada associação nacional tem a responsabilidade pela elaboração detalhada de suas próprias especificações de competência. O NCB (National Competence Baseline) em coerência com o ICB (IPMA Competence Baseline) e a cultura local. As especificações nacionais podem incluir novos elementos, detalhar ou simplificar os elementos já descritos no ICB, de maneira razoável (20% dos itens), de modo a contemplar particularidades da cultura local. No Brasil, existe o Brazilian National Competence Baseline.

No Brasil, a ABGP (Associação Brasileira em Gerenciamento de Projetos), é a única instituição associada que representa a IPMA. Sua "cartilha" é o Referencial Brasileiro de Competências em Gerenciamento de Projetos (da mesma forma que o PMI tem o seu o corpo de conhecimento, o PMBoK). Porém, o RBC, não contempla apenas os elementos de conhecimento do gerenciamento de projetos, mas outras competências do profissional de gerenciamento de projetos. Contém 37 elementos:

1- Projetos e gerenciamento de projetos;

2- implementação do gerenciamento de projetos;

3- gerenciamento por projetos;

4- abordagem sistêmica e integração;

5- contexto do projeto;

6- fases e ciclos de vida do projeto;

7- desenvolvimento e avaliação do projeto;

8- objetivos e estratégias do projeto;

9- critérios de sucesso e insucesso do projeto;

10- iniciação do projeto;

11- encerramento do projeto;

12- estruturas do projeto;

13- conteúdo e escopo;

14- programação do tempo;

15- recursos;

16- custos e financiamento do projeto;

17- configuração e modificações;

18- riscos do projeto;

19- medida do desempenho;

20- controle do projeto;

21- informação, documentação e reporting;

22- organização do projeto;

23- trabalho em equipe;

24- liderança;

25- comunicação;

26- conflitos e crises;

27- aquisições e contratos;

28- qualidade do projeto;

29- informática em projetos (icb 29);

30- normas e regulamentações (icb 30);

31- negociação e reuniões (icb 32);

32- marketing e gerenciamento de produtos (icb 38);

33- segurança, saúde e meio ambiente (icb 40)

34- aspectos legais (icb 41);

35- gestão da cadeia de suprimentos;

36- trabalho colaborativo à distância;

37- gestão do conhecimento em projetos.

## 2.1.6 Gerenciamento de riscos

### 2.1.6.1 Conceito

Processo sistemático de definição, análise e resposta aos riscos do Projeto.

### 2.1.6.2 Objetivo

Maximizar a probabilidade de ocorrência e impacto dos eventos considerados positivos (oportunidades) e minimizar no caso de eventos negativos (ameaças).

### 2.1.6.3 Principais processos da gerência de riscos do projeto

#### a. Planejamento do gerenciamento de riscos

No planejamento são tomadas as decisões sobre como abordar, planejar e executar as atividades de gerenciamento de riscos do *Projeto*. Os riscos se manifestam por vários motivos, alguns são internos e outros externos. O ambiente do Projeto, o planejamento, os processos de gerenciamento, os recursos inadequados, podem contribuir nesse sentido. Alguns riscos serão conhecidos antecipadamente, outros ocorrerão de forma inesperada durante o Projeto. É importante saber qual é o nível de tolerância a riscos da organização e dos *stakeholders* antes de avaliá-los e ordená-los.

> O *Guide to the PMBOK* afirma que o Planejamento do Gerenciamento de Riscos é um fundamento crucial de todos os processos que se seguem, no sentido de "(...) assegurar que o nível, tipo e visibilidade do gerenciamento de riscos estejam de acordo com o risco e a importância do projeto em relação à organização, para fornecer tempo e recursos suficientes para as atividades de gerenciamento de riscos e para estabelecer *uma base acordada de avaliação de riscos*" (HELDMAN, 2006. p.163).

#### b. Identificação de riscos

A determinação dos riscos que podem afetar o Projeto e a documentação de suas características é um processo interativo, que passa por constantes renovações. Trata-se também de investigação, para obter informações confiáveis através de ferramentas e técnicas. Além de identificar os riscos, é preciso identificar os sintomas que possam ser monitorados para permitir uma atitude pro ativa perante a eventual proximidade de ocorrência do risco.

### c. Análise qualitativa de riscos

É o processo de avaliar o impacto e a probabilidade dos riscos identificados. É a priorização dos riscos para análise ou ação subseqüente por meio de avaliação e combinação de sua probabilidade de ocorrência e impacto.

Permite qualificar e classificar os riscos em função do seu efeito potencial individual e priorizá-los em função do seu efeito potencial para o projeto como um todo. Impactos potenciais sobre os objetivos e probabilidade de ocorrência dos riscos são avaliados segundo categorias que expressam a graduação de intensidade e o nível de tolerância que servem de referência comum.

Categorias como Muito Alto ("MA"), Alto ("A"), Médio ("M"), Baixo ("B") ou Muito Baixo ("MB") são definidas segundo critérios objetivos.

### d. Análise quantitativa de riscos

São os processos de medição da probabilidade e do impacto dos riscos e estimativa suas implicações nos objetivos do Projeto.

É o processo que avalia os impactos e quantifica a exposição do Projeto aos riscos por meio da atribuição de probabilidades numéricas a cada um e aos seus impactos sobre os objetivos do Projeto.

### e. Planejamento de respostas aos riscos

Desenvolvimento de opções e ações para aumentar as oportunidades e reduzir as ameaças aos objetivos do Projeto.

Ele especifica as medidas a serem tomadas para reduzir ameaças e tirar proveito das oportunidades encontradas nos processos de análise de riscos.

É a gravidade do risco que determina o nível de planejamento das respostas a ser feito. As respostas devem ser compensadoras em termos de custo.

**f. Monitoramento e controle dos riscos**

O monitoramento e controle de riscos deve se dar ao longo de todo o ciclo de vida do Projeto, monitorando os riscos identificados e reexaminando os planos para averiguar se serão capazes de lidar com os riscos de maneira adequada diante de sua aproximação durante esse processo.

Envolve, primeiramente, a identificação e resposta aos riscos assim que ocorrem.

O registro de riscos deve ser atualizado em duas circunstâncias: a primeira é quando a reavaliação ou auditoria de riscos conclui que algum elemento da informação original sobre o risco foi alterado. A segunda se dá quando o risco precisa ser encerrado

Quando se tem uma ocorrência de um evento de risco, ela será documentada no registro de riscos juntamente com a eficácia de seu plano de resposta.

**2.1.6.4 O PMBOK trata o assunto de Gerência de Riscos, através de 6 etapas**

a. *Planejamento da Gerência de Riscos*, onde se toma a decisão sobre como esta vai ser abordada e como serão planejadas as atividades de gerência de riscos.

**b.** *Identificação dos Riscos*, onde se procura identificar todos os riscos, positivos ou negativos para seu projeto. A descrição do Risco deve ser completa e autocontida, procurando-se no texto da descrição do risco se contemplar Causa e Conseqüência – qual o risco e o que ele provoca.

**c.** *Análise Qualitativa dos Riscos*, onde se procura dimensionar categorias para o risco, tipo Muito Alto ("MA"), Alto ("A"), Médio ("M"), Baixo ("B") ou Muito Baixo ("MB"), visando criar um processo para se selecionar os riscos que deverão ser tratados.

**d.** *Análise Quantitativa dos Riscos*, onde se procura determinar a Probabilidade e o Impacto dos riscos no projeto, de forma a que possamos comparar riscos de diversos tipos entre si.

**e.** *Planejamento de Respostas aos Riscos*, onde decidimos o que vamos fazer para responder a cada Risco. Isto inclui para os riscos negativos, Aceitar o Risco, Mitigar (diminuir seus efeitos) ou Eliminar, e para os Riscos Positivos, Ignorar (não fazer nada), Melhorar (tentar aumentar seus efeitos) ou Provocar (tentar fazer acontecer).

**f.** *Controle dos Riscos*, onde vamos acompanhar o que foi planejado. Esta atividade é especialmente importante, pois os Riscos podem mudar – morrerem riscos identificados, aumentar ou diminuir a probabilidade ou o impacto, ou até surgirem novos riscos.

## 2.2 Modelos de maturidade em gerência em projetos

Modelos de maturidade em gerenciamento de projetos vêm obtendo notoriedade e diversas organizações, entidades normativas, pesquisadores e consultores organizacionais têm desenvolvido normas e modelos de referência que buscam promover o desenvolvimento das competências em gerenciamento de projetos.

## 2.2.1 Capability Maturity Model® Integration

O Capability Maturity Model Integration (CMMI) é um modelo de referência que contém práticas necessárias à maturidade em disciplinas específicas (Systems Engineering (SE), Software Engineering (SE), Integrated Product and Process Development (IPPD), Supplier Sourcing (SS)). Desenvolvido pelo SEI (Software Engineering Institute), o CMMI é uma evolução do CMM e procura estabelecer um modelo único para o processo de melhoria corporativo, integrando diferentes modelos e disciplinas em um único programa de melhoria facilitando a padronização, estruturação, termos utilizados e forma de medição da maturidade. (FIG. 1)

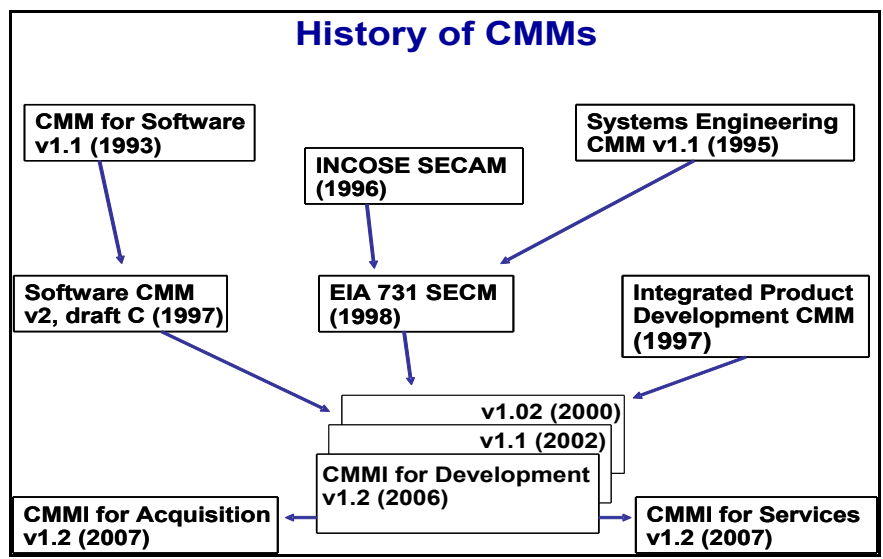

**Figura 1:** The History of CMMs.

**Fonte:** *SEI CMMI for Development Version 1.2.* 2006. p. 07

Os processos de melhoria nasceram de estudos realizados por Deming (1986) (*Out of the Crisis*), Crosby (1979) (*Quality is Free: The Art of Making Quality*

*Certain*) e Juran (1945) (*Management of inspection and quality control*), cujo objetivo principal era a melhoria da capacidade dos processos. Entende-se por capacidade de um processo a habilidade com que este alcança o resultado desejado.

Um modelo tem como objetivo estabelecer - com base em estudos, históricos e conhecimento operacional - um conjunto de "melhores práticas" que devem ser utilizadas para um fim específico.

O CMMI objetiva, além da integração dos modelos e redução dos custos com melhorias de processo, os seguintes objetivos também fazem parte do projeto CMMI:

- Aumento do foco das atividades;
- Integração dos processos existentes;
- Eliminar inconsistências;
- Reduzir duplicações;
- Fornecer terminologia comum;
- Assegurar consistência com a norma ISO 15504;
- Flexibilidade e Extensão para outras disciplinas.

## 2.2.1.1 "O que" x "como"

O modelo não descreve processo algum, são orientações definidas através das práticas especificadas. O processo específico é estabelecido e descrito pela empresa usuária do modelo, são, portanto, políticas e procedimentos internos interligados entre si operando dentro do CMMI. (FIG. 2)

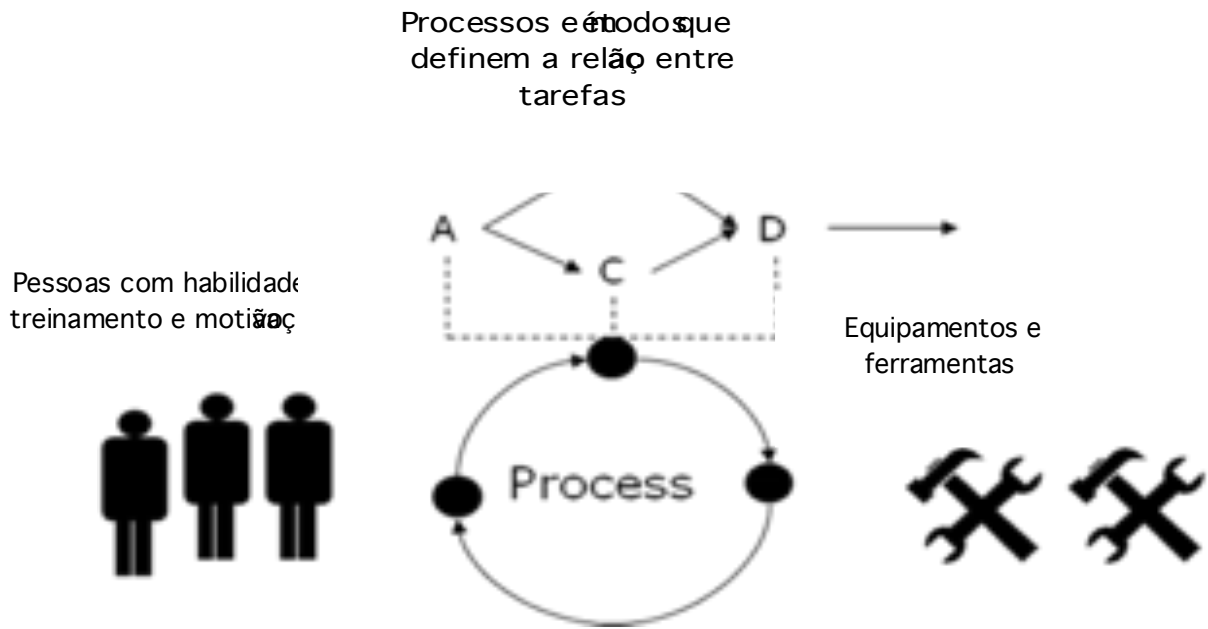

**Figura 2:** As três dimensões críticas.

**Fonte:** *SEI CMMI for Development Version 1.2. 2006. p. 04*

Os componentes do modelo são os mesmos para a representação contínua e por estágios. São eles: áreas de processo, metas específicas, práticas específicas, metas genéricas, práticas genéricas, produtos de trabalho típicos, sub-práticas, amplificações de disciplinas, elaboração de praticas genéricas e referências.

## 2.2.1.2 Representações

O CMMI possui duas representações distintas definidas como contínua e por estágios. O uso de uma ou outra representação é uma escolha aberta a organização. O uso de ambas também é possível na solução de necessidades particulares dentro dos programas de melhoria uma vez que cada forma de representação possui vantagens uma sobre a outra.

## 2.2.1.1.1 Contínua

A representação contínua tem como resultado uma mapa de Níveis de Capacidade extraído por avaliação de agrupamento das Áreas de Processo por Categoria. (FIG. 3)

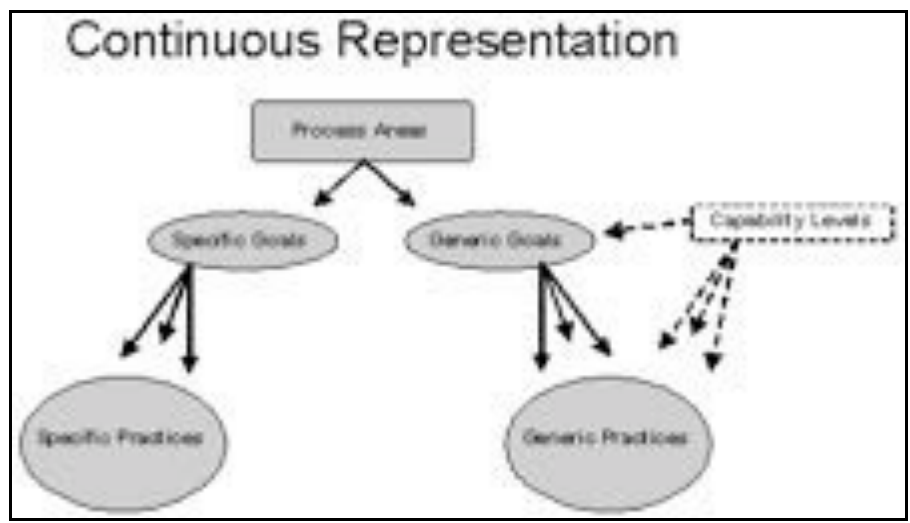

**Figura 3:** Estrutura da Representação Contínua.

**Fonte:** *SEI CMMI for Development  Version 1.2.  2006. p. 30*

A representação contínua tem como vantagens os seguintes aspectos:

• Fornece maior flexibilidade focando em áreas de processo específicas de acordo com metas e objetivos de negócio uma vez que a organização pode estabelecer metas individuais para cada área de processo direcionando assim seus recursos para áreas críticas ou de maior interesse desta.

• Permite a comparação de áreas de processo entre diferentes organizações. Quando as áreas processos são agrupadas de forma semelhante entre organizações é possível estabelecer um *benchmarking* entre estas áreas.

- Estrutura familiar para aqueles que estão migrando da comunidade de engenharia de sistemas.

- Foco bem definido nos riscos específicos de cada área de processo.

- Estrutura compatível com o padrão ISO/IEC 15504.

- Um nível de capacidade é um plano bem definido que descreve a capacidade de uma área de processo.

- Cada nível representa uma camada na base para a melhoria contínua do processo.

- Assim, níveis de capacidade são cumulativos, ou seja, um nível de capacidade mais alto inclui os atributos dos níveis mais baixos.

**2.2.1.1.1.1   Existem seis níveis de capacidade de 0 a 5 definidos como:**

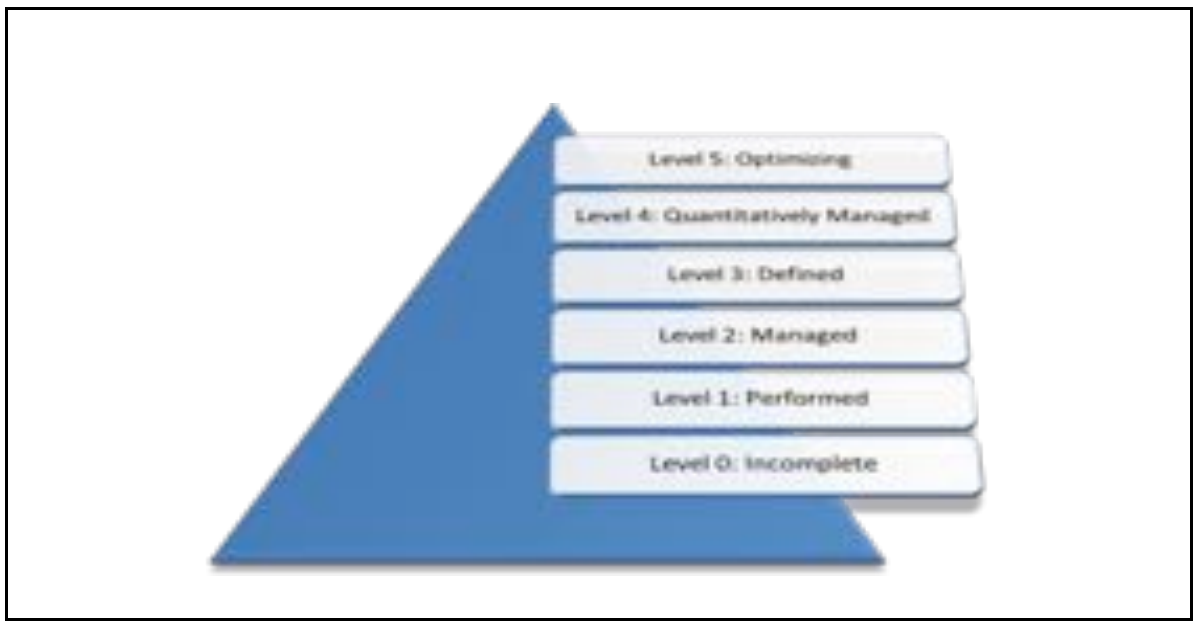

**Figura 4:** Os seis níveis de capacidade.

**Fonte:** ANTUNES, Ricardo Magno Acervo próprio.

- **Capacidade Nível 0: Incompleto**

Um "processo incompleto" é um processo que não é executado ou parcialmente não executado. Um ou mais metas específicas da área do processo não é satisfeito, e nenhum objetivo genérico existe para este nível uma vez que não há razão para institucionalizar um processo parcialmente executado.

- **Capacidade Nível 1: Executado**

Um processo executado é um processo que satisfaz as metas específicas de uma área de processo. Isto suporta e habilita o trabalho necessário para a realização de produtos de trabalho.

Embora capacidade nível 1 possua melhorias importantes,estas melhorias podem se perder ao longo do tempo caso não sejam padronizadas e institucionalizadas.

- **Capacidade Nível 2: Gerenciado**

Um processo gerenciado é executado (capacidade nível 1) e possui infraestrutura básica implantada para suportar este processo. O processo é planejado e executado de acordo com políticas, com pessoas capazes e habilitadas e com recursos adequados para produzir resultados controlados; envolve interessados relevantes; é monitorado, controlado, revisado; e é avaliada quanto à aderência a descrição do processo.

- **Capacidade Nível 3: Definido**

Um processo definido é um processo gerenciado (capacidade nível 2) que é entrelaçado a organização através do estabelecimento de padrões de processo, e contribui para o produto do trabalho, medições, e outras melhorias para o ativo de processos organizacionais. A descrição dos processos no nível 3 é mais rigorosa que a descrição realizada no nível 2.

- **Capacidade Nível 4: Gerenciado Quantitativamente**

Um projeto gerenciado quantitativamente é um processo definido (capacidade nível 3) que é controlado utilizando estatística e outras técnicas quantitativas. Objetos quantitativos para qualidade e desempenho de processos são estabelecidos e utilizados como critérios de gerenciamento. Qualidade e desempenho de processo é entendida em termos estatísticos e gerenciados durante a vida do processo.

- **Capacidade Nível 5: Otimizado**

Um processo otimizado é qualitativamente gerenciado (capacidade nível 4) que é melhorado baseado no entendimento de causas comum das variações inerentes no processo.O foco de uma otimização do processo é continuamente aperfeiçoar a variação do processo em ambos incremento e inovação.

- **Avançando através dos níveis de capacidade**

Os níveis de capacidade de uma área de processos são alcançados através da aplicação de práticas genéricas ou alternativas apropriadas a processos associados com a determinada área de processo. (FIG. 5)

Alcançar o nível de capacidade 1 para uma área de processos equivale a dizer que os processos associados a esta área são executados.

O nível de capacidade 2 é atingido quando existe uma política, conforme citado no item Capacidade Nível 2, para execução da área de processos.

A padronização da área de processos padronizado na organização indica o que o nível de capacidade 3 foi alcançado.

Ao atingir o nível de capacidade 4 assume-se que a área de processo é gerenciada utilizando técnicas estatísticas e quantitativas.

O nível de capacidade 5 assume que os sub-processos estão estabilizados e reduções em causas comuns de variações destes processos são reduzidas.

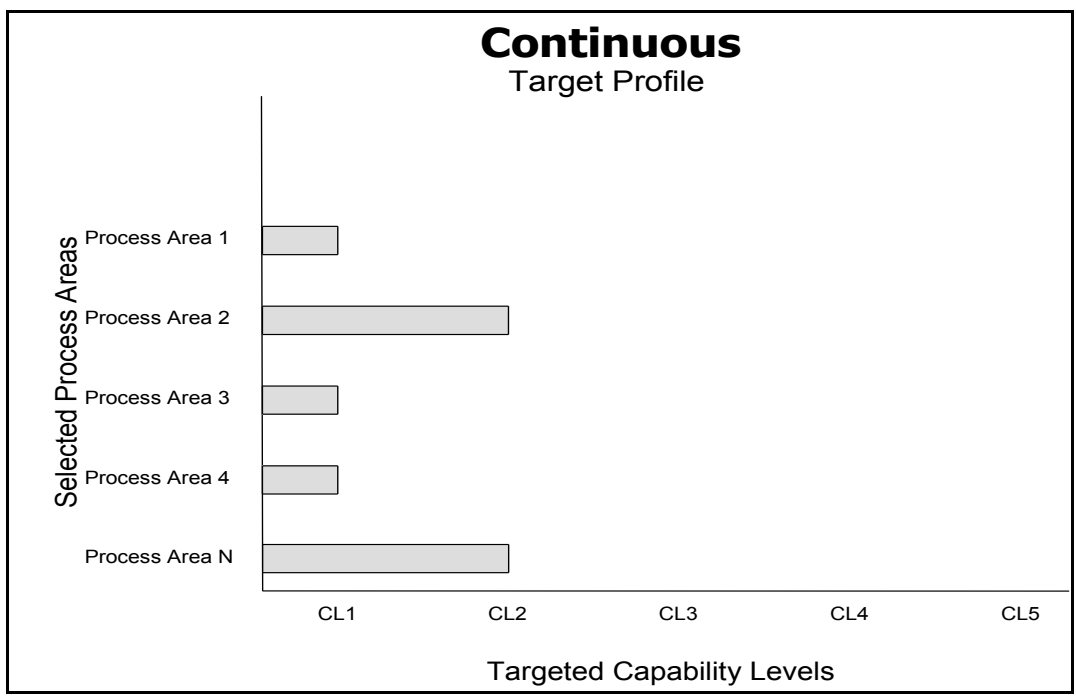

**Figura 5:** Alvo dos Níveis de Capacidade.

**Fonte:** *SEI CMMI for Development Version 1.2.* 2006. p. 41

- **Por Estágios**

A representação por estágios tem como resultado uma foto do Nível de Maturidade Organização como um todo, extraído por avaliação de agrupamento das Áreas de Processo por Nível. (FIG. 6)

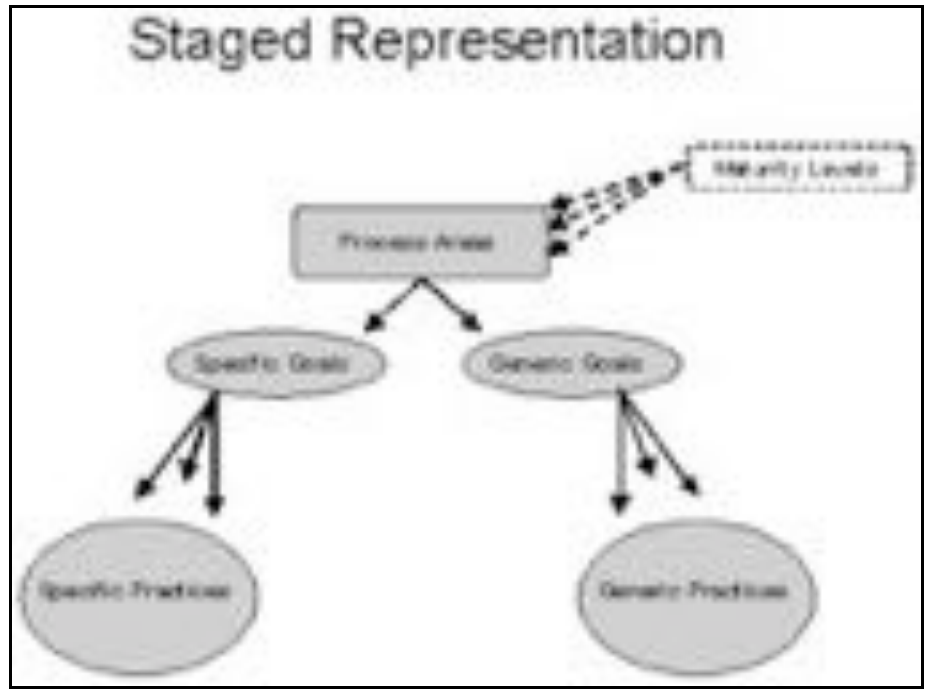

**Figura 6:** Estrutura da Representação por Estágios.

**Fonte:** *SEI CMMI for Development Version 1.2.* 2006. p. 30

- **As vantagens da representação por estágios são as seguintes:**

Fornece uma rota de implementação bem definida através de grupos de área de processo, implementação seqüencial e melhoria (cada nível funciona como a fundação para o próximo nível). Similar ao PDCA (O ciclo PDCA, ciclo de Shewhart ou ciclo de Deming, foi introduzido no Japão após a guerra, idealizado por Shewhart (1939) e divulgado por Deming, quem efetivamente o aplicou. O ciclo de Deming tem por princípio tornar mais claros e ágeis os processos envolvidos na execução da gestão, como por exemplo, na gestão da qualidade, dividindo-a em quatro principais passos. Plan (planejamento): estabelecer missão, visão, objetivos (metas), procedimentos e processos (metodologias) necessários para atingir os resultados. Do (execução): realizar, executar as atividades. Check (verificação): monitorar e avaliar periodicamente os resultados, avaliar processos e resultados, confrontando-os com o planejado, objetivos, especificações e estado desejado, consolidando as informações, eventualmente confeccionando relatórios. Act (ação): Agir de acordo com o avaliado e de acordo com os relatórios, eventualmente determinar e confeccionar novos planos de ação, de forma a melhorar a qualidade, eficiência e eficácia, aprimorando a execução e corrigindo eventuais falhas.)

Possui estrutura familiar para aqueles que estão migrando do SW-CMM

Habilidade de gerenciar processos através da organização. Atribui uma nota de classificação do nível de maturidade em que a organização se encontra através dos resultados das avaliações. Permitindo dessa forma a comparação de forma direta entre as organizações.

## 2.2.1.1.1.2 Existem 4 níveis de maturidade de 1 a 4 definidos como:

- **Maturidade nível 1: inicial**

Na maturidade nível 1, os processos são usualmente caóticos e sem que nenhuma técnica reconhecida empregada e/ou cujas fases variam em cada aplicação do processo. A organização geralmente na prove um ambiente estável ao suporte do processo. O sucesso nas organizações provém da capacidade de indivíduos e não da execução de processos de eficácia comprovada. Organizações de maturidade nível 1 produzem freqüentemente excedem orçamentos e prazos embora entreguem produtos e serviços com qualidade variável.

- **Maturidade nível 2: gerenciado**

A maturidade nível 2 possui processos planejados e executados de acordo com políticas, com pessoas capazes e habilitadas e com recursos adequados para produzir resultados controlados; envolve interessados relevantes; é monitorado, controlado, revisado; e é avaliado quanto a aderência a descrição do processo.

Os processos neste nível são repedidos e suportados por um plano documentado, o estado dos produtos e entregas de serviços é visível a gerência em pontos definidos (exemplo, grandes marcos e encerramento de atividades importantes). Os produtos e serviços entregues são controlados apropriadamente e atendem satisfatoriamente a descrições específicas do processo, padrões e procedimentos.

- **Maturidade nível 3: definido**

Os processos na maturidade nível 3 são bem caracterizados e entendidos, e são descritos em padrões, procedimentos, ferramentas e métodos. A organização define os processos, os estabelece e aprimora durante o tempo. Esses processos estabelecem consistência pela organização. A descrição dos processos no nível 3 é mais rigorosa que a descrição realizada no nível 2. Processos na maturidade nível 3 são gerenciados pro-ativamente utilizando o entendimento das relações entre atividades e medições dos processos, produtos e serviços.

- **Maturidade nível 4: gerenciada quantitativamente**

Neste nível a organização e os projetos estabelecem metas quantitativas para a qualidade de desempenho dos processos assim como critérios de gerenciamento desses. As metas são baseadas nas necessidades de clientes, usuários finais, da organização e daqueles que implementam os processos. Qualidade e desempenho de processos são entendidos em termos estatísticos e gerenciados através de seu ciclo de vida (SEI, 2001).

- **Maturidade nível 5: otimizado**

A organização, neste nível é capaz de continuamente aprimorar os processos envolvidos baseado no entendimento de causas comuns de variações inerentes aos processos. *Maturity level 5 focuses on continually improving process performance through incremental and innovative process and technological improvements.* Processos aprimoram-se através de metas quantitativas, continuamente revisadas, estabelecidas pela organização refletindo os objetivos desta. (FIG. 7; FIG. 8)

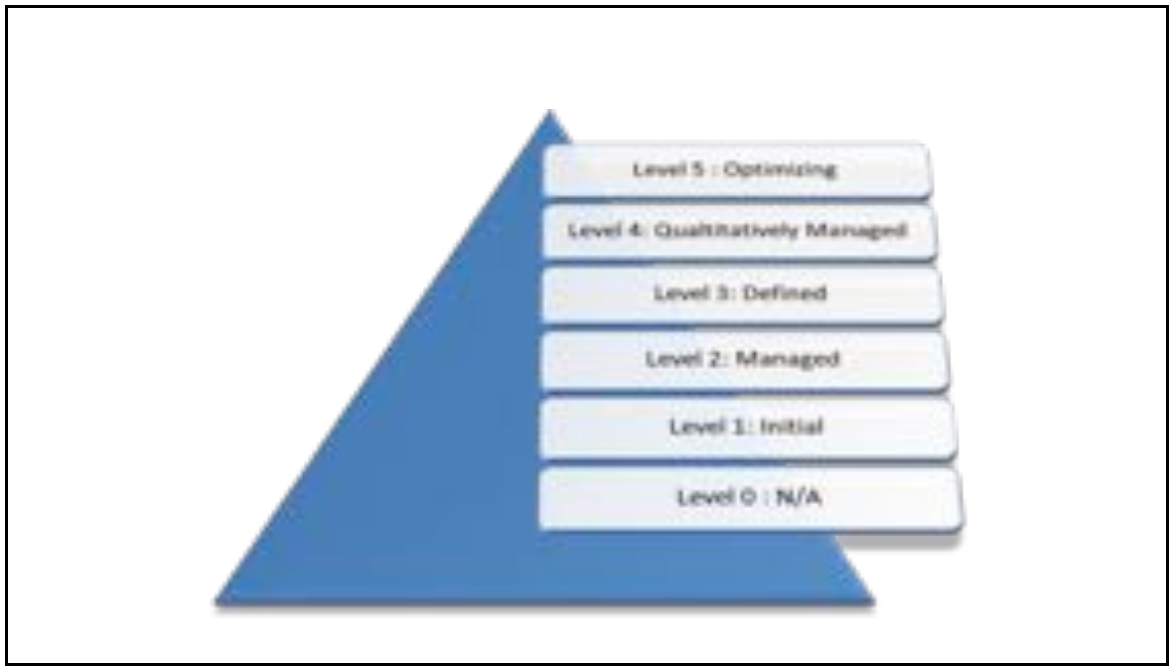

**Figura 7:** Níveis de Maturidade.

**Fonte:** ANTUNES, Ricardo Magno  Acervo próprio.

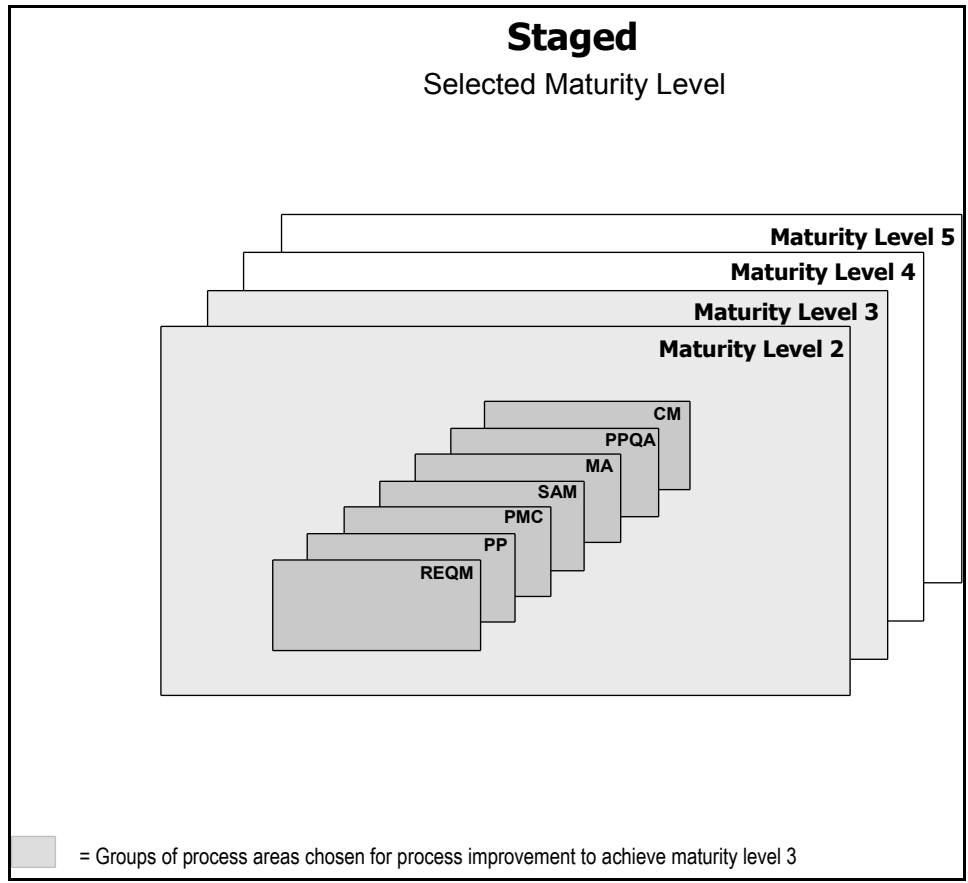

**Figura 8:** Seleção do Nìvel de Maturidade.

**Fonte:** *SEI CMMI for Development Version 1.2.* 2006. p. 41

- **Avaliações do CMMI**

O método de avaliação utilizado *Standard* CMMI Assessment Method for Process Improvement (SCAMPI) também desenvolvido pelo Software Engineering Institute (SEI) e são descritos no Standard CMMI Appraisal Method for Process Improvement SM (SCAMPI), Version 1.2: Method Definition, versão atual até a data deste trabalho. A avaliação é realizada por entidades externas a organização e autorizados pelo SEI como avaliadores. Existem cerca de 180 avaliadores SCAMPI no mundo.

## 2.2.2 *ProjectFRAMEWORK™*

### 2.2.2.1 Contextualização

Em 1999, o *ESI International* lançou o seu modelo de maturidade aplicado à gestão de projetos – *ProjectFRAMEWORK™*.

No *ProjectFRAMEWORK™* foram incorporadas as idéias de progressividade e de níveis de maturidade do modelo CMMSM, cujo campo de aplicação restringe-se às organizações que desenvolvam softwares. (FIG. 9)

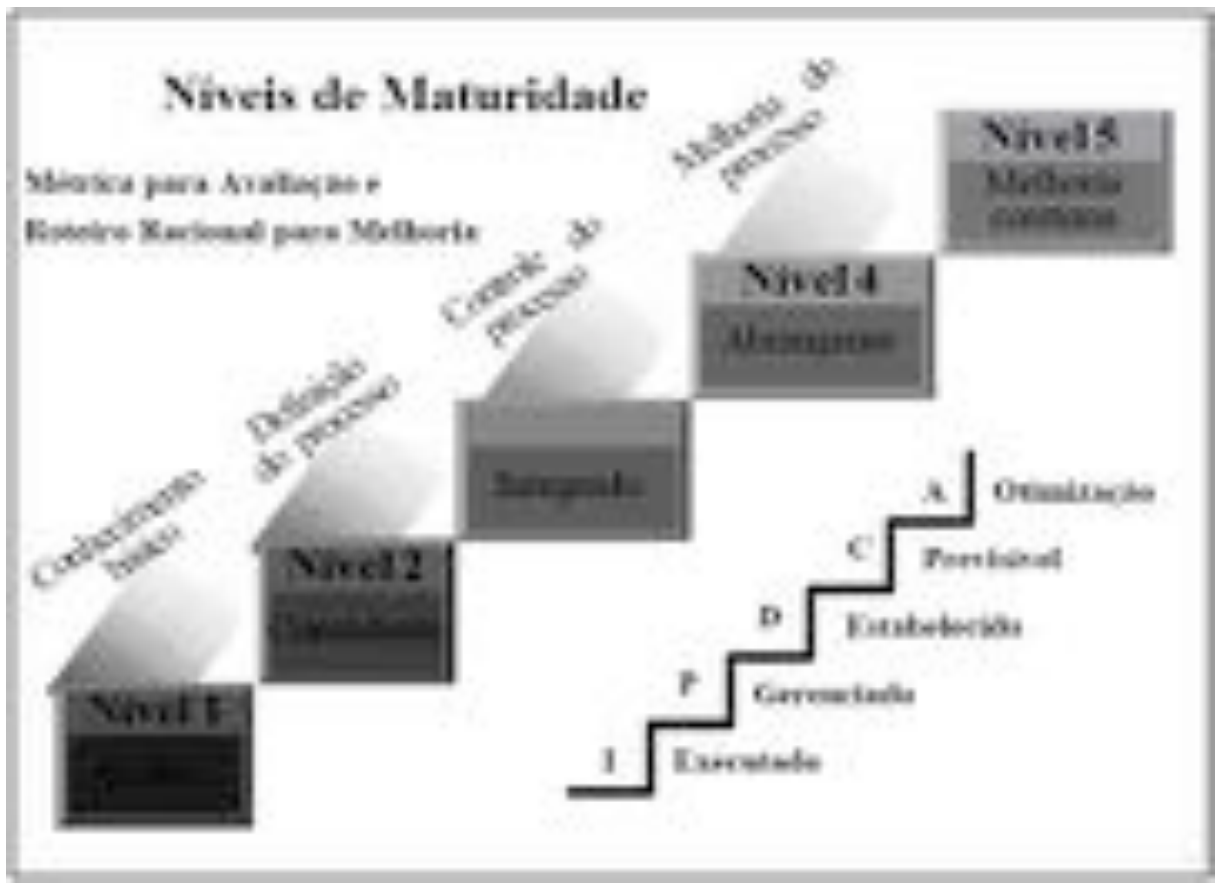

**Figura 9:** Níveis de maturidade-métrica para avaliação e roteiro racional para melhoria.

**Fonte:** OLIVEIRA, W. L. *Visão geral do projectframework*. 2002. Slide 4. www.simpros.com.br.

Ao mesmo tempo em que o *ProjectFRAMEWORK™* apresenta objetivos de desempenho para cada uma das nove áreas de conhecimento do PMBOK® Guide, visando o aperfeiçoamento contínuo da gestão de projetos nas organizações através da integração de pessoas, processos e tecnologia, ele também permite diagnosticar adequadamente os processos de gestão de projetos e reduzir o escopo de melhorias a serem introduzidas no plano de ação resultante, objetivando passar de um nível ou estágio de maturidade para o nível superior.

O *ProjectFRAMEWORK™* propõe apoiar as organizações:

- Identificando pontos fortes e fracos no processo de gestão de projetos;

- Estabelecendo uma referência na capacitação em gerência de projetos;

- Tornando-se projetizados com resultados previsíveis;

- Lançando um programa de melhoria contínua;

- Integrando princípios e processos de gestão eficaz de projetos na estrutura e nos processos organizacionais.

## 2.2.2.2 Quais são os níveis?

(1) Ad Hoc, (2) Consistente, (3) Integrado, (4) Compreensivo e (5) Otimizado.

## 2.2.2.3 Como são definidos? Como passamos de um nível para outro?

O nível inicial do modelo é chamado *"Ad hoc"*, pois, os processos de gestão de projetos, se houver, são mal definidos e geralmente aplicados nos projetos independentemente um do outro. A repetição dos métodos e processos não pode ser claramente discernida, dado que há pouco ou nenhum suporte na organização para gestão eficaz de projetos.

O segundo nível chamado "Consistente" é alcançado quando o amadurecimento já levou a organização a apoiar uma abordagem disciplinada na execução dos processos básicos empregados na gestão dos seus projetos. As responsabilidades e papeis são identificados. Os recursos que irão executar são recrutados, treinados e alocados para realizar cada processo seguindo políticas e diretrizes bem definidas. Os processos são documentados e repetidos em toda a organização. Os projetos são gerenciados da mesma maneira por diferentes gerentes de projeto. Os processos são revisados pela alta administração e ações corretivas são tomadas sistematicamente. Uma metodologia de gestão de projetos é introduzida, com apoio da direção da organização.

A organização passa para o terceiro nível, conhecida dentro do modelo como "Integrado", quando os processos de gestão de projetos forem ajustados para avançar e melhorar aspectos específicos das nove áreas de conhecimento, bem como os cinco processos definidos pelo PMBOK® Guide. O alinhamento desses processos, observando as melhores práticas das áreas de conhecimento, estabelece a fundação de conhecimento e utilização de uma metodologia integrada em toda a organização.

O amadurecimento do quarto nível "Compreensivo" é alcançado quando a metodologia da gestão de projetos já foi totalmente implementada em toda a organização. Ela é efetivamente usada para medir a eficácia e reduzir as variações na execução de projetos. As técnicas avançadas e ferramentas alternativas são desenvolvidas e usadas. Os interessados, internos e externos, são encarados como parceiros no processo de gestão e os projetos dão suporte e são ligados ao plano estratégico da organização.

No último estágio "otimizado" a organização procura continuamente por inovações em busca de excelência em gestão de projetos. As causas comuns de falhas dos projetos são eliminadas.

Além dos níveis acima descritos, o modelo é composto de áreas de conhecimento com focos específicos em função do nível de maturidade, de objetivos para cada área de conhecimentos e do desempenho. Este último quesito é desdobrado em: (a) compromisso com o desempenho, (b) habilidade para desempenhar e (c) desempenho real. O modelo se completa com um sistema de medição.

## 2.2.2.4 Comparação entre os níveis de maturidade 1 e 2 (formação de base) na gerência dos riscos do projeto

| Ad hoc | Consistente |
|---|---|
| ▪ Inexiste um processo de gerenciamento de riscos formal;<br>▪ Riscos só são reconhecidos quando transformam-se em problemas;<br>▪ A organização não promove ou encoraja a equipe a discutir riscos. | ▪ A organização reconhece a importância do gerenciamento dos riscos dos projetos;<br>▪ Riscos são identificados e analisados, e respostas aos riscos são planejadas;<br>▪ A ênfase principal são riscos que envolvem impactos no cronograma e custo;<br>▪ Riscos geralmente são analisados individualmente sem considerar múltiplos efeitos. |

## 2.2.2.5 Base do gerenciamento de projetos na organização – nível "consistente"

O nível "Consistente" de maturidade é caracterizado por um processo metodológico, o apoio gerencial e o treinamento.

### 2.2.2.5.1 Processo metodológico

- É desenvolvida e introduzida uma metodologia de gerenciamento de projetos.

- Um processo repetível é estabelecido para o gerenciamento de projetos.

- Os gerentes de projeto preparam planos de projeto, estimam e controlam o desempenho dos projetos de acordo com um processo estabelecido.

- As capacidades estabelecidas nesse nível possibilitam as organizações. eliminar muitos problemas relacionados ao cumprimento de prazos, custos e objetivos de qualidade dos seus projetos.

### 2.2.2.5.2 Apoio gerencial

- A gerência executiva da organização claramente apóia as políticas e compromissos para o desenvolvimento de capacidades em gerenciamento de projetos.

- Políticas estabelecidas promovem um senso de responsabilidade e disciplina no desenvolvimento dos processos de gerenciamento de projetos.

- Quando estes processos são institucionalizados, a organização tem uma base para construir métodos aperfeiçoados e processos.

- Responsabilidades são atribuídas para funções específicas de gerentes de projetos e os profissionais desenvolvem essas funções com treinamentos específicos.

### 2.2.2.5.3 Treinamento

- As habilidades do nível 2 são apoiadas quando gerentes de projetos, membros da equipe e gerentes seniores da organização são identificados e completam programas completos de treinamento em gerenciamento de projetos.

- Cada treinamento garante que as pessoas na organização têm o conhecimento, habilidade e competência requerida para desenvolver suas atribuições.

- Os treinamentos incluem a metodologia de gerenciamento de projetos, *software* de gerenciamento de projetos, formação de equipes, análise de requisitos, gerenciamento de custos, identificação e quantificação de riscos e gerenciamento de aquisições.

### 2.2.2.6 Conclusão

Os níveis 1, 2 e 3 são analisados em termos de todas as nove áreas de conhecimento onde cada área tem seu próprio objetivo de melhorar o processo de gestão de projetos.

No nível 4, o foco é direcionado às três áreas requeridas para institucionalizar a gestão de projetos na organização: Gestão de Integração; Gestão de Recursos Humanos e Gestão de Riscos.

No último e quinto nível, toda a atenção é voltada para aprimorar a gestão da integração.

Os objetivos são definidos, em cada área de conhecimento, visando descrever o que deve acontecer nesta área para assegurar que as melhores práticas sejam aplicadas. O indicador da capacidade organizacional é o alcance dos objetivos traçados, em cada nível de maturidade. Em outras palavras, os objetivos significam o escopo, as fronteiras e o intento de cada área de conhecimento.

Finalmente, a medição se faz necessária para determinar se os pré-requisitos foram respeitados e as atividades foram institucionalizadas. O sistema de medição deve conter dois componentes que permitam a avaliação, via coleta e análise das métricas, e a verificação da aderência às políticas e às diretrizes estabelecidas na gestão de projetos, através de revisões e auditorias periódicas. (FIG. 10)

**Figura 10:** Modelo de maturidade – *ProjectFRAMEWORK™*.
**Fonte:** IETEC. Boletim Gestão de Projetos no. 13. 2004. Disponível em: www.ietec.com.br.

### 2.2.3 Prado - MMGP V4

### 2.2.3.1 Contextualização

O modelo Prado-MMGP foi desenvolvido entre 1998 e 2002 e publicado em Dezembro de 2002. Ele baseia-se em níveis e reflete a experiência do autor na implantação de gerenciamento de projetos em organizações brasileiras. Ele deve ser aplicado separadamente a cada setor da organização. (FIG. 11)

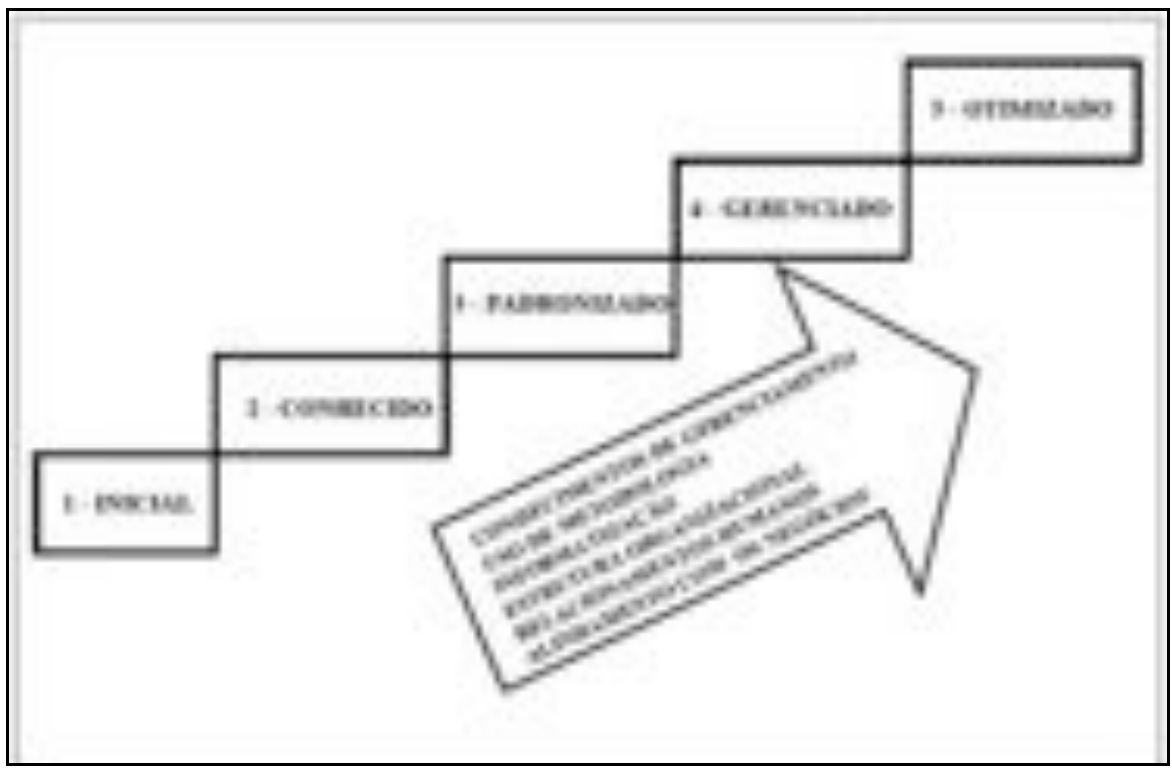

**Figura 11:** Dimensões e níveis de maturidade.

**Fonte:** PRADO, D. *Um modelo brasileiro de maturidade em GP.* 2006. Slide 14.

### 2.2.3.2 Dimensões da maturidade setorial

As dimensões a seguir estão presentes em cada nível de maturidade. O diferencial é determinado pelo momento em que ocorre o pico de maturidade em uma determinada dimensão.

- Conhecimentos de gerenciamento;

- Uso prático de metodologia;

- Informatização;

- Estrutura organizacional;

- Relacionamentos humanos;

- Alinhamento com os negócios da organização.

### 2.2.3.3 Níveis de maturidade

O modelo Prado-MMGP apresenta 5 níveis de maturidade, conforme Figura 2. A evolução nos níveis ocorre segundo 6 dimensões. (FIG. 12)

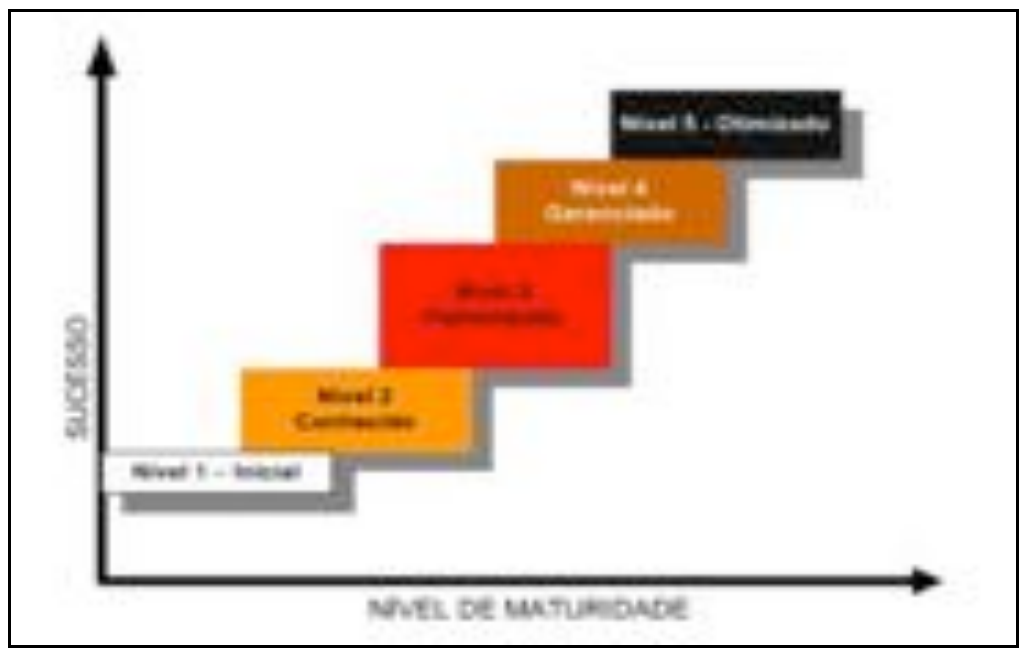

**Figura 12:** Níveis de Maturidade modelo Prado-MMGP.

**Fonte:** PRADO, D. *Visão resumida do Modelo Prado – MMGP.* 2007. p. 1

Os níveis de maturidade são:

- Inicial.

- Conhecido (Linguagem Comum).

- Padronizado.

- Gerenciado.

- Otimizado.

**QUADRO 1**

**Dimensões da maturidade x níveis de maturidade**

| Dimensão da Maturidade | Nível de Maturidade | | | | |
|---|---|---|---|---|---|
| | 1 Inicial | 2 Conhecido | 3 Padronizado | 4 Gerenciado | 5 Otimizado |
| Conhecimentos | Dispersos | Básicos | Básicos | Avançados | Avançados |
| Metodologia | Não há | Tentativas Isoladas | Implantada e Padronizada | Estabilizada | Otimizada |
| Informatização | Tentativas Isoladas | Tentativas Isoladas | Implantada | Estabilizada | Otimizada |
| Estrutura Organizacional | Não há | Não há | Implantada | Estabilizada | Otimizada |
| Relacionamentos humanos | Boa vontade | Algum avanço | Algum avanço | Algum avanço | Maduros |
| Alinhamento com estratégias | Não há | Não há | Iniciado | Alinhado | Alinhado |

**QUADRO 2**

**Características de cada nível de maturidade**

# QUADRO 3

## Resumo descritivo dos níveis de maturidade

| Descrição dos Níveis de Maturidade do Modelo Prado-MMGP (Setorial) | |
|---|---|
| NÍVEL | DESCRIÇÃO |
| 1 | **Inicial ou Embrionário ou ad hoc** : a empresa está no estágio inicial de gerenciamento de projetos, que são executados na base da "boa vontade" ou do "melhor esforço" individual. Geralmente não se faz planejamento e o controle é inexistente. Não existem procedimentos padronizados. O sucesso fruto do esforço individual ou da sorte. As possibilidades de atraso, estouro de orçamento e não atendimento às especificações técnicas são grandes. |
| 2 | **Conhecido** : A organização fez investimentos constantes em treinamento e adquiriu *softwares* de gerenciamento de projetos. Pode ocorrer a existência de iniciativas isoladas de padronização de procedimentos, mas seu uso é restrito. Percebe-se melhor a necessidade de se efetuar planejamento e controle e, em algumas iniciativas isoladas, alguma melhoria é percebida. No restante os fracassos "teimam" em continuar ocorrendo. |
| 3 | **Definido ou padronizado** : foi feita uma padronização de procedimentos, difundida e utilizada em todos os projetos sob a liderança de um Escritório de Gerenciamento de Projetos (EGP). Uma metodologia está disponível e é praticada por todos e parte dela está informatizada. Foi implementada uma estrutura organizacional adequada e possível ao setor e aos seus tipos de projetos no momento da implementação. Tenta-se obter o melhor comprometimento possível dos principais envolvidos. Os processos de planejamento e controle são consistentes e o processo de aprendizagem faz que eles sejam executados cada vez melhor.Os resultados "estão aparecendo". |
| 4 | **Gerenciado** : Os processos estão consolidados e a empresa está aperfeiçoando o modelo através da coleta e da análise de um banco de dados sobre projetos executados. Ele possibilita uma avaliação da causa de desvios da meta dos projetos e contramedidas estão sendo estabelecidas e aplicadas. O Ciclo de Melhoria Continua é aplicado sempre que se detecta alguma deficiência. A estrutura organizacional é revista e evolui para outra que permite um relacionamento mais eficaz com as áreas envolvidas (eventualmente uma estrutura projetizada, matricial balanceada ou forte). Existe um alinhamento dos projetos com os negócios da organização. Os gerentes estão se aperfeiçoando ainda mais em aspectos críticos do gerenciamento, tais como relacionamentos humanos, conflitos, negociações, etc. A aplicação de processos de gerenciamento de projetos é reconhecida como fator de sucesso para os projetos. |
| 5 | **Otimizado** : Existe uma otimização na execução de projetos com base na larga experiência e também nos conhecimentos e atitudes pessoais (disciplina, liderança, etc). Os novos projetos podem também se basear em um excelente banco de dados de "melhores práticas". O nível de sucesso é próximo de 100%. A organização tem alta confiança em seus profissionais e aceita desafios de alto risco. |

## 2.2.3.4 Avaliação da maturidade

Através do questionário MMGP será obtido tanto o valor global da maturidade quanto o perfil de aderência aos diversos níveis.

O conceito "Percentual de Aderência" deve ser aplicado conjuntamente com o nível de maturidade para se perceber melhor o estágio de maturidade de uma organização. Desta forma o valor obtido no Teste de Avaliação de Maturidade reflete quão bem a organização se posiciona nos requisitos daquele nível.

Os valores obtidos para cada nível serão classificados da seguinte forma:

- Aderência até 20%: nula ou fraca;

- Aderência de 20% até 60%: Regular;

- Aderência de 60% até 90%: Boa;

- Aderência acima de 90%: Completa.

É utilizada para estabelecer um "Plano de Ação" tanto a média obtida quanto o percentual de aderência.

### 2.2.3.5 Conclusão

O modelo de maturidade Prado-MMGP mostra claramente que a evolução no assunto deve ocorrer em diversas dimensões e que, depois de implantada uma dimensão, se gasta algum tempo para que se possa classificar como madura aquela dimensão. A seqüência tem-se mostrado um conveniente "Mapa de Raciocínio" para muitas organizações.

### 2.2.4 Project Management Maturity Model (PMMM)

Kerzner (2001) descreve a maturidade em GP como o desenvolvimento de metodologias e processos que são naturalmente repetitivos e garantem uma alta probabilidade que cada projeto atinja o sucesso. A partir do nível 3, forma um ciclo contínuo e repetitivo pelo qual a organização atinge a excelência em GP. De acordo com o modelo Kerzner de maturidade de projetos a organização deve resolver 183 questões de múltipla escolha referentes aos cinco níveis de maturidade. (FIG. 13)

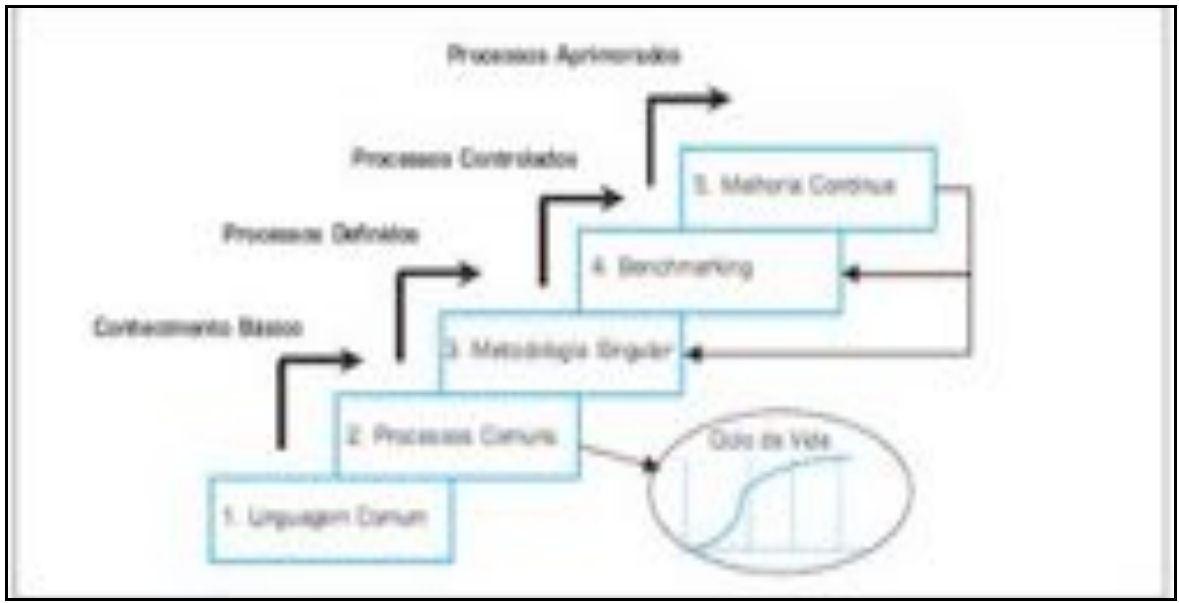

**Figura 13:** Project Management Maturity Model (PMMM).

**Fonte:** Adaptado de KERNER, 2001. p. 42

### a. Nível 1: Linguagem comum (80 questões)

Neste nível de maturidade a organização reconhece a importância do gerenciamento de projetos. O nível 1 é baseado no conhecimento dos princípios fundamentais do gerenciamento de projetos e de sua terminologia. O nível 1 pode ser completado através de um bom entendimento do PMBOK. Características:

- Uso esporádico da metodologia de gerenciamento de projetos.
- Nenhum suporte do nível executivo da organização.
- Nenhum investimento em treinamento e educação em gerenciamento de projetos.
- Nenhuma ação que transpareçam o reconhecimento dos benefícios do gerenciamento de projetos.

**b. Nível 2: processos comuns (20 questões)**

Neste nível de maturidade a organização faz um esforço para usar gerenciamento de projetos e desenvolver processos e metodologias para suportar seu uso efetivo. A organização percebe que metodologias e processos comuns são necessários de forma que o sucesso de gerencia de um projeto possa ser replicado a outros. Características:

- Reconhecimento dos benefícios do gerenciamento de projetos.
- Suporte organizacional para todos os níveis.
- Reconhecimento da necessidade de um controle de custos.
- Desenvolvimento de um currículo de treinamento em gerenciamento de projetos.

**c. Nível 3: Metodologia única (42 questões)**

Neste nível de maturidade a organização reconhece que o sinergismo e controle de processo podem ser atingidos mais facilmente através de uma metodologia única ao invés de usar múltiplas metodologias. Companhias que atingiram o nível 3 são totalmente comprometidas ao conceito Gerenciamento de Projetos. Características:

- Processos integrados.
- Suporte cultural.
- Suporte de gerenciamento em todos os níveis.
- Gerenciamento de projetos informal.
- Obtenção de retorno do investimento de treinamentos em GP.
- Excelência comportamental.

### d. Nível 4: Benchmarking (25 questões)

Neste nível de maturidade a organização usa *benchmarking* para continuamente comparar práticas de gerenciamento de projetos de lideres reconhecidos para conseguir informações para ajudá-los a melhorar sua *performance*. *Benchmarking* é um esforço contínuo de analise e avaliação. Os fatores críticos de sucesso para o *benchmarking* são geralmente os processos chave do negocio e como eles são integrados. Características:

- Estabelecimento de um escritório de gerenciamento de projetos(PMO) ou de um centro de excelência.
- Dedicação ao processo de *benchmarking*.

### e. Nível 5: Melhoria contínua (16 questions)

Neste nível de maturidade, a organização avalia a informação aprendida durante o benchmarking e implementa as mudanças necessárias para melhorar o processo de gerenciamento de projetos. A organização percebe que excelência em gerenciamento de projetos é uma jornada sem fim.

- Transferência de conhecimento.
- Programa de *mentoring* por parte do PMO.
- Realização de planejamento estratégico para gerenciamento de projetos.

## 2.2.5 P3M3

O modelo da maturidade da gerência de projeto (P3M3) pode ser usado como base para melhorar os processos da gerência do *portfólio*, do programa e de projeto.

É estruturado com cinco níveis da maturidade. O modelo de maturidade em gerência de projeto (P3M3) é uma versão realçada do modelo de maturidade em gerência de projeto, baseada no processo de maturidade do *framework* que evoluiu através do instituto de engenharia de *software* (SEI) *Capability Maturity Model* (CMM). Entretanto, desde que P3M3 foi projetado a SEI revisou radicalmente seus modelos de maturidade para criar CMMI.

A experiência de SEI entre 1986-91 indicou que os questionários da maturidade fornecem uma ferramenta simples para identificar as áreas onde os processos de uma organização podem necessitar de melhoria.

Concluiu-se que o desenvolvimento de um modelo descritivo da referência seria em fornecer as organizações benefícios dando direções mais eficazes para estabelecer programas de melhoria de processo.

O questionário não é a base para melhorar processos da gerência do programa e de projeto, o ponto chave é o modelo.

Os níveis descritos dentro do P3M3 indicam como as áreas chaves do processo podem ser estruturadas hierarquicamente para fornecer estados da transição para uma organização que deseja ajustar objetivos realísticos e sensíveis para a melhoria.

Os níveis facilitam transição organizacional de um estado imaturo para transformar-se em uma organização madura e capaz com uma base objetiva para julgar a qualidade e resolver edições do programa e do projeto.

O ponto chave é que o modelo, não um questionário é a base para melhorar processos da gerência do programa e de projeto.

Uma organização madura tem uma habilidade de organização para os programas e os projetos controlando baseados no programa e em processos padrão, definidos pela gerência de projeto.

O modelo da maturidade da gerência de projeto (P3M3) (QUADRO 4) pode ser usado como a base melhorando processos da gerência do *portfolio*, do programa e de projeto. É estruturado com cinco níveis da maturidade, que são:

- Nível 1 - processo inicial.
- Nível 2 - processo repetível.
- Nível 3 - processo definido.
- Nível 4 - processo controlado.
- Nível 5 – processo otimizado.

**QUADRO 4**

**Níveis da Maturidade P3M3**

| Maturity: | Project | Programme | Portfolio |
|---|---|---|---|
| **Level 1 - initial process** | Does the organisation recognise projects and run them differently from its ongoing business? (Projects may be run informally with no standard process or tracking system.) | Does the organisation recognise programmes and run them differently to projects? (Programmes may be run informally with no standard process or tracking system.) | Does the organisation's Board recognise programmes and projects and run an informal list of its investments in programmes and projects? (There may be no formal tracking and reporting process.) |
| **Level 2 - repeatable process** | Does the organisation ensure that each project is run with its own processes and procedures to a minimum specified standard? (There may be limited consistency or co-ordination between projects) | Does the organisation ensure that each programme is run with its own processes and procedures to a minimum specified standard? (There may be limited consistency or co-ordination between programmes) | Does the organisation ensure that each programme and/or project in its portfolio is run with its own processes and procedures to a minimum specified standard? (There may be limited consistency or co-ordination) |
| **Level 3 - defined process** | Does the organisation have its own centrally controlled project processes, **and** can individual projects flex within these processes to suit the particular project? | Does the organisation have its own centrally controlled programme processes **and** can individual programmes flex within these processes to suit the particular programme? | Does the organisation have its own centrally controlled programme and project processes **and** can individual programmes and projects flex within these processes to suit particular programmes and/or projects. And does the organisation have its own portfolio management process? |
| **Level 4 - managed process** | Does the organisation obtain and retain specific measurements on its project management performance **and** run a quality management organisation to better predict future performance? | Does the organisation obtain and retain specific measurements on its programme management performance **and** run a quality management organisation to better predict future programme outcomes? | Does the organisation obtain and retain specific management metrics on its whole portfolio of programmes and projects as a means of predicting future performance? Does the organisation assess its capacity to manage programmes and projects and prioritise them accordingly? |
| **Level 5 - optimised process** | Does the organisation run continuous process improvement **with** proactive problem and technology management for projects in order to improve its ability to depict performance over time and optimise processes? | Does the organisation run continuous process improvement **with** proactive problem and technology management for programmes in order to improve its ability to depict performance over time and optimise processes? | Does the organisation run continuous process improvement **with** proactive problem and technology management for the portfolio in order to improve its ability to depict performance over time and optimise processes? |

**2.2.6 P2MM**

O P2MM é muito similar ao modelo da maturidade da gerência do *portfolio*, do programa e de projeto do (P3M3), mas consiste em somente três níveis.

P2MM permite uma organização de medir seu nível de perícia em gerência de projeto utilizando PRINCE2. O modelo da maturidade PRINCE2 (P2MM) avalia somente a maturidade nos primeiros três destes níveis e para a gerência de projeto somente. Conjuntamente com o P3M3, permite uma organização executar boa prática em encontro a uma marca de nível de maturidade reconhecida.

O modelo da maturidade (P2MM) desde que o lançamento de PRINCE2 em 1996, as organizações esforçaram-se continuamente para executar eficazmente o método.

A finalidade do modelo da maturidade PRINCE2 é permitir organizações de calibrar, pela avaliação, sua maturidade no uso do método da gerência de projeto. O modelo pode também ser usado:

- Para compreender as práticas chave que é parte de um processo organizacional eficaz para controlar projetos.

- Para identificar as práticas da chave que necessitam ser encaixadas dentro da organização para conseguir o nível seguinte da maturidade.

- Para compreender a rotatividade atrás do questionário da avaliação.

Pode ser usado de uma ou duas maneiras:

- Como um modelo autônomo da maturidade.

- Como um subconjunto da modalidade mais larga da maturidade da gerência do *portfolio*, do programa e de projeto de OGC.

O modelo da maturidade PRINCE2 pode ser usado por qualquer um:

- Como um modelo autônomo da maturidade para que as organizações avaliem o nível da maturidade para seu uso de PRINCE2.

- Conjuntamente com o *portfolio* da OGC, modelo da maturidade da gerência do programa e de projeto (P3M3), para as organizações que adotaram o uso de PRINCE2, para avaliar o nível da maturidade para a gerência de seus projetos.

O nível 1 é o nível baixo para ambos os modelos. É provável que uma organização neste nível em um modelo estará no mesmo nível em ambos. Entretanto, é possível que:

- Uma organização poderia estar em um nível mais elevado no P1M3 e estar neste nível no P2MM, porque somente começou apenas adotar o uso de PRINCE2

- Uma organização poderia estar em um nível mais elevado no P2MM e estar neste nível no P1M3, porque o P3M3 requer uma cobertura mais larga de disciplinas da gerência de projeto no nível 2.

O nível 2 é amplamente equivalente porque PRINCE2 fornece o mínimo especificou o padrão requerido neste nível pelo P1M3. Entretanto, PRINCE2 não se dirige especificamente à gerência da parte interessada e do fornecedor, que são áreas chaves do P1M3.

O nível 3 tem diferenças mais significativas neste nível porque PRINCE2 não se dirige especificamente ao seguinte:

- Gerência de informação.
- A escala cheia do treinamento e da instrução de gerência do projeto.
- Gerência da matriz através da organização.

Os níveis 4 e 5, coordenação e *networking* do Inter-grupo, estes níveis no P1M3 têm um foco forte sobre:

- Medida do desempenho do processo do projeto.
- Melhoria contínua.

Nenhum destes é dirigida por PRINCE2.

O formato de cada área do processo usa três títulos padrão:

- Área Chave do Processo.
- Finalidade.
- A chave pratica destes títulos fornece uma estrutura que descreve o que uma organização deve fazer para estabelecer e encaixar cada processo.

Estes são derivados da publicação "projetos bem sucedidos controlando de OGC com PRINCE2", que contém a cobertura cheia de cada uma das áreas do processo.

A estrutura hierárquica nos níveis descritos dentro do modelo da maturidade PRINCE2 indica como as áreas do processo chaves podem ser estruturas hierárquicas para fornecer estados da transição para uma organização que deseja ajustar objetivos realísticos e sensibilidade para a melhoria.

Os níveis facilitam a melhoria organizacional de um nível relativamente baixo da potencialidade do processo para transformar-se em uma organização madura e capaz com uma base objetiva para julgar a qualidade do projeto e resolver edições do projeto.

Toda a terminologia usada dentro deste modelo é consistente com a aquela usada dentro do manual PRINCE2. Onde a referência é feita a um produto particular da gerência PRINCE2 (por exemplo, relatório do destaque), isto deve conformar ao esboço do produto no apêndice A do PRINCE2 manual ou a uma descrição do produto, que seja produzida para substituir o esboço do produto. (FIG. 14)

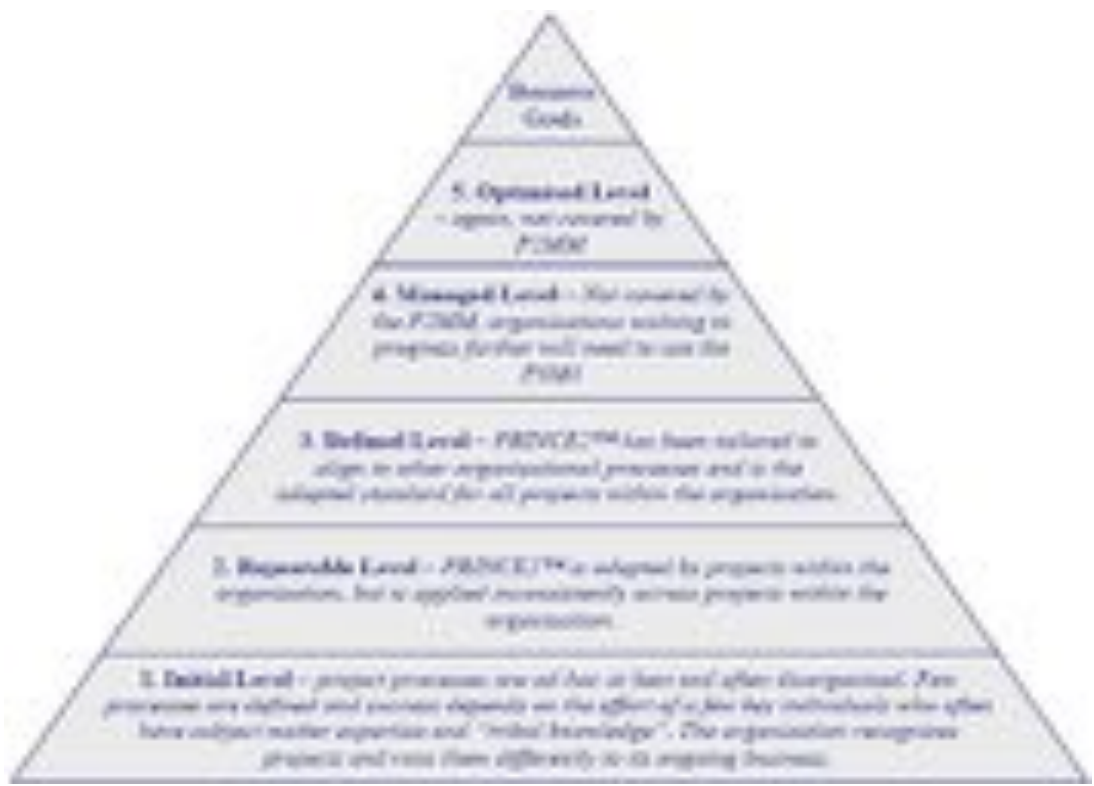

**Figura 14:** Níveis da maturidade P2MM.

**Fonte:** *ILX Group plc.* 2006

**2.2.7 OPM3**

Uma ferramenta que tem se mostrado bastante eficiente é o OPM3®. Esta é a sigla para *"Organizacional Project Management Maturity Model",* um produto desenvolvido pelo próprio PMI® e lançado no mercado no final de 2003.

Neste método usa-se um *software* que de acordo com o resultado de 151 perguntas gera quatro gráficos e listas das melhores práticas em gerenciamento de projetos ou da maturidade em gerenciamento de projeto (bem como Programas e *portfólios*). (FIG. 15)

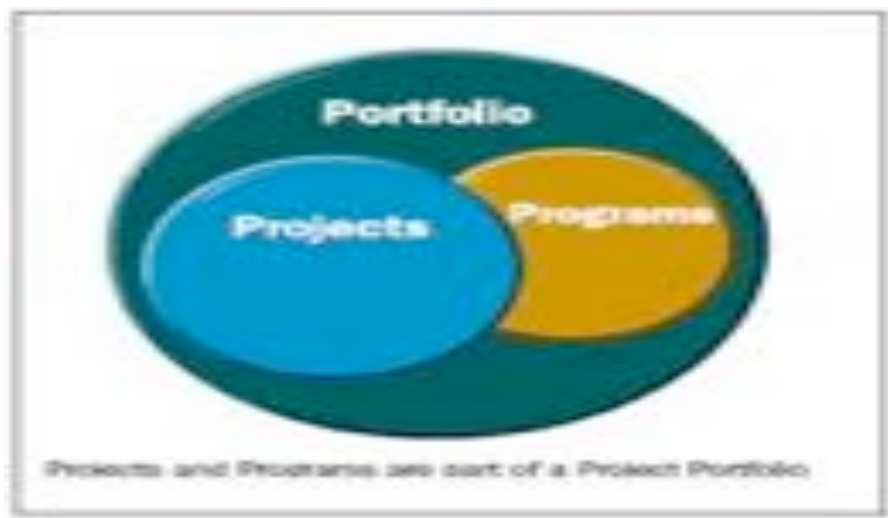

**Figura 15: Projetos e programas são partes de um portfólio de projetos.**

**Fonte: PROJECT MANAGEMENT INSTITUTE, INC.** *Organizational Project Management Maturity Model (OPM3®) Knowledge Foundation* **2003. p. 20**

Este resultado permite visualizar a situação atual da Gestão de Projetos, levando em consideração uma série de atributos. São eles: a maturidade em Projetos, Programas e *Portfólios* e o grau de utilização das melhores práticas em quatro outros atributos: Padrões, Métricas, Controles e Melhorias. (FIG. 16)

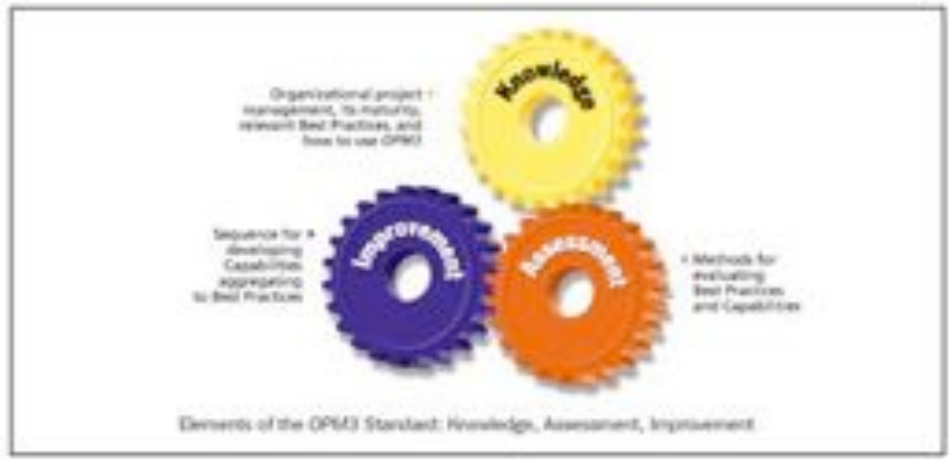

**Figura 16: Conhecimento direciona a avaliação, e ela direciona as melhorias**

**Fonte: PROJECT MANAGEMENT INSTITUTE, INC.** *Organizational Project Management Maturity Model (OPM3®) Knowledge Foundation* **2003. p. 15**

Permite também traçar um objetivo a ser alcançado em números absolutos para as mesmas variáveis, levando em consideração os objetivos da empresa, o tipo de negócio e também um *benchmarking* de empresas similares. Entender a diferença entre o valor absoluto de maturidade "atual" de uma empresa e o "desejado" permite o desenvolvimento de um plano de longo prazo permitindo estimar o esforço e o tempo necessário para implementar as melhorias. (FIG. 17)

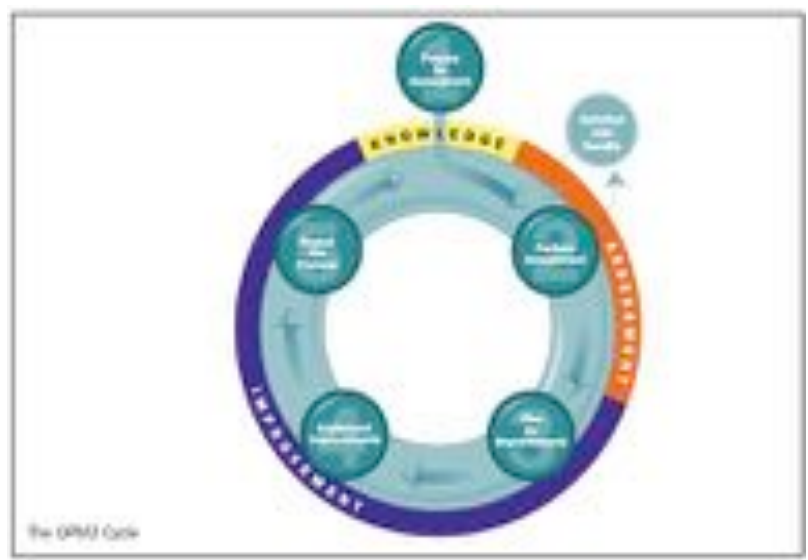

**Figura 17: Ciclo OPM3.**

**Fonte: PROJECT MANAGEMENT INSTITUTE, INC.** *Organizational Project Management Maturity Model (OPM3®) Knowledge Foundation* **2003. p. 25**

Normalmente, os resultados são aplicados na solução de algumas situações:

## 2.2.7.1 Visibilidade da evolução

Mais comumente, esta evolução é mostrada sem que se tenha uma "foto" da situação atual. Muitas frentes de melhoria e capacitação, quando aplicadas simultaneamente, sem uma avaliação da situação atual, podem ocasionar uma tendência do time a perder a noção dos processos que foram aplicados inicialmente e cair na síndrome do "eu sei que não gerencio os projetos corretamente". Esta síndrome se reflete negativamente em muitos pontos da equipe, normalmente em uma baixa alto-estima que quanto mais evolui, mais tem a sensação de precisar evoluir. Outras conseqüências dessa síndrome são: dimensionar "folgas" de planejamento sempre mais generosas que o necessário e na falta de confiança na execução de processos. (FIG. 18)

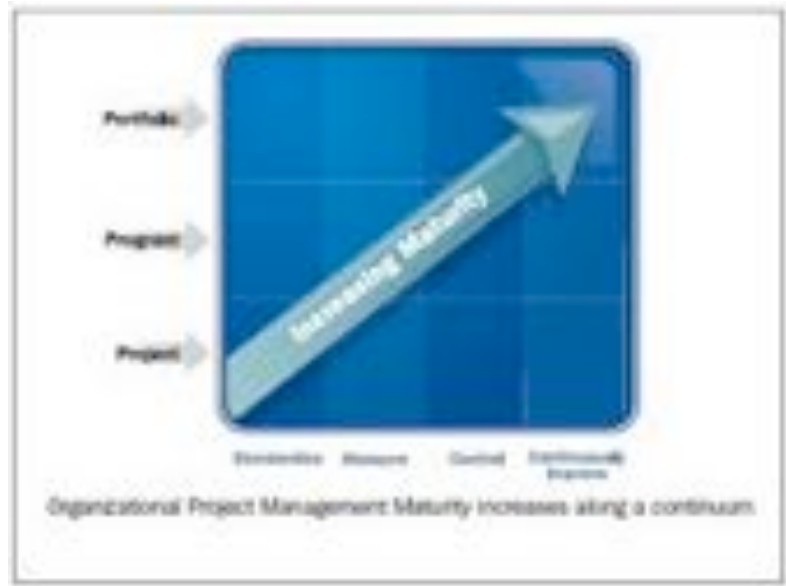

**Figura 18: A Maturidade em Gerenciamento de Projetos cresce continuamente**

Fonte: PROJECT MANAGEMENT INSTITUTE, INC. *Organizational Project Management Maturity Model (OPM3®) Knowledge Foundation* 2003. p. 22

**2.2.7.2 Visão pragmática de um plano de evolução de maturidade**

Um resultado bastante importante da análise é o de trazer mais segurança, embasamento e realismo para um plano de evolução de maturidade em gestão de projetos.

**2.2.7.3 Excelência operacional e nivelamento de padrões**

A definição de um padrão ou "gabarito" para uma determinada área ou linha de negócio permite através do OPM3®, verificar quais as melhores práticas que precisam ser desenvolvidas e desenvolver para cada uma das unidades de negócio um plano realista de tempo, necessidades de infra-estrutura, padrões e capacitação. O que na prática traz bastante velocidade e flexibilidade na implementação de mudanças.

**2.2.7.4 Quantificar o custo da mudança**

Um resultado bastante interessante também tem vindo da utilização em avaliação de empresas para aquisição, pois, permite estimar um valor financeiro para adequação processual da empresa a ser adquirida. Esta informação permite planejar o investimento necessário para se ter uma operação executando conforme o planejado.

Em suma, existem várias aplicações da análise de maturidade da gestão de projetos que incrementam o negócio e que são convertidas em resultado direto para o negócio. Uma das demandas que o diagnóstico busca atender é a necessidade de desenvolver um direcionamento de como atuar nesta etapa de mudança cultural de forma bastante simples e objetiva.

# 3 METODOLOGIA CIENTÍFICA

## 3.1 Considerações iniciais

Este capítulo descreve o método utilizado neste trabalho e buscou-se avaliar a estrutura contida em modelos de maturidade e capacidade existentes no mercado. Foram escolhidos os que apresentam maior aderência de mercado e de diferentes estruturas e aplicações a fim de se extrair os melhores critérios para a criação do modelo de maturidade em gerenciamento de riscos proposto neste trabalho.

Os critérios a serem analisados dos modelos citado são os seguintes:

• **Aderência e/ou vinculo a metodologia específica:**

O modelo mede o grau de maturidade em uso de alguma metodologia ou prática descrita especificamente ou esse mede de formal geral a maturidade da área de processos.

• **Aderência ao gerenciamento de riscos:**

O modelo em algum momento cita a gerência de riscos e possui critérios para mensurar a organização neste aspecto.

• **Aplicabilidade:**

Existe alguma restrição a aplicação do modelo em uma organização, isto é, o modelo deve ser implantado em organizações de tamanho negocio ou outro critério definido.

• **Seqüenciamento:**

Qual a lógica para se implantar o modelo caso essa exista.

• **Escalonabilidade:**

Como modelo deve ser implantando, ou partes convenientes à organização podem ser utilizadas enquanto as que não são tão relevantes são deixadas de fora sem esta fragmentação afetar o resultado final da avaliação do modelo.

• **Número de níveis de capacidade/maturidade:**

Como o modelo divide os diferentes estágios de maturidade e como os classifica.

• **Tipo de métrica utilizado:**

Qual a unidade de medida para a composição do nível de maturidade.

• **Critérios para mudança de níveis:**

Quando o avaliado muda de nível, isto é, qual os critérios para a evolução do nível de maturidade da organização.

•**Forma de avaliação:**

Como o modelo, ou avaliadores que utilizam o modelo extraem os resultados de maturidades da organização frente aos critérios exigidos pelo modelo.

• **Forma de apresentação de resultados:**

Forma em que o modelo apresenta o resultado final ao avaliado.

• **Complexidade de aplicação:**

A aplicação do modelo exige cuidados especiais como consultorias ou auditorias, estas podem ser internas ou externas. Existem alguma certificação ou autorização necessária aos avaliadores por órgãos credenciados.

## QUADRO 5

**Comparativo entre os níveis de capacidade/maturidade dos modelos citados**

| Nível | CMMI Contínuo | CMMI Por estágios | P2MM | PMMM | P3M3 | Project Framework | Prado MMGP | OPM3 |
|---|---|---|---|---|---|---|---|---|
| 0 | Incompleto | N/A | N/A | N/A | N/A | N/A | N/A | N/A |
| 1 | Executado | Inicial | Inicial | Linguagem Comum | Inicial | Ad Hoc | Inicial | N/A |
| 2 | Gerenciado | Gerenciado | Repetível | Processos Comuns | Repetível | Consistente | Conhecido | Standardize |
| 3 | Definido | Definido | Definido | Metodologia Singular | Definido | Integrado | Padronizado | Measure |
| 4 | Quantitativamente Gerenciado | Quantitativamente Gerenciado | N/A | Benchmarking | Gerenciado | Compreensivo | Gerenciado | Control |
| 5 | Otimizado | Otimizado | N/A | Melhoria Contínua | Otimizado | Otimizado | Otimizado | Continuosly Improve |

## 3.2 O modelo de maturidade em gerenciamento de riscos em projetos

O modelo proposto visa atender as práticas comuns de gerenciamento de riscos em projetos executadas pelas instituições citadas anteriormente além de outras, sem um vínculo formal com qualquer metodologia ou guia de práticas.

Medir a aderência as melhores práticas de gerenciamento de riscos é o principal objetivo deste modelo embora não seja o único aspecto levado em conta. Fatores que disseminam e entrelaçam o gerenciamento de riscos ao negocio da organização são fatores importantes no resultado obtido da aplicação do modelo.

O modelo deve ser aplicado a projetos ou organizações que aplicam projetos como principal negócio. Tarefas repetitivas ou empreendimentos não caracterizados como projetos podem obter resultados incorretos ou influenciados de alguma forma pois o modelo visa mensurar o gerenciamento de riscos em projetos utilizando práticas desta área. O tamanho do projeto é irrestrito uma vez que são levadas em consideração práticas gerais.

Existe um seqüenciamento de avaliação que segue o ciclo básico do gerenciamento de riscos em projetos.

Conforme o gerenciamento de riscos o modelo deve ser implantado ou avaliado em sua totalidade a fim de avaliar todo o ciclo do gerenciamento de riscos. A fragmentação do modelo influencia de modo negativo o resultado, tendenciando-o para baixo uma vez que itens de avaliação não obterão nota.

Dividimos o modelo em cindo níveis de capacidade/maturidade baseado em conceitos comuns em outros modelos, possibilitando a comparação em termos gerais com demais modelos. A escolha dos cinco níveis também está baseada nas escalas de Thurstone ou Guttman e Likert como métodos de escalonamento unidimensional. Definido o foco de medir o nível de maturidade em gerenciamento de riscos em projetos, objetivo único por se tratar de um método unidimensional de escalonamento, foram adotados os cincos níveis de Likert embora os graus de aceite bipolar, isto é níveis positivos e negativos de aceite, a determinada proposição foram abandonas e utilizamos a polaridade única do método Guttman, isto é, menos favorável a mais favorável ao conceito. Embora sejam cinco níveis de menor a maior aderência as melhores práticas em gerenciamento de riscos contempladas no modelo de maturidade proposto. Os cinco níveis, em ordem numérica crescente de menor a maior aderência foram denominadas: Nível 1 – Inicial; Nível 2 – Definido; Nível 3 – Gerenciado; Nível 4 – Gerenciado Quantitativamente; Nível 5 – Otimizado. (FIG. 19)

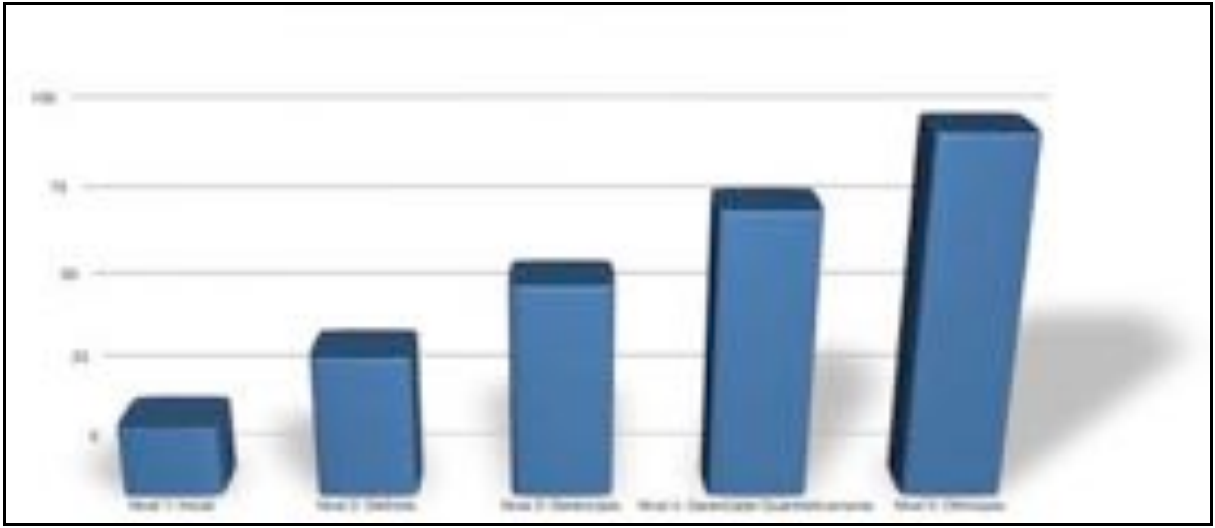

**Figura 19:** Níveis de maturidade em gerenciamento de riscos em projetos.

**Fonte:** ANTUNES, Ricardo Magno Acervo próprio.

**a. Tipo de métrica utilizado:** Qual a unidade de medida para a composição do nível de maturidade. Critérios para mudança de níveis. Quando o avaliado muda de nível, isto é, quais os critérios para a evolução do nível de maturidade da organização. (FIG. 20)

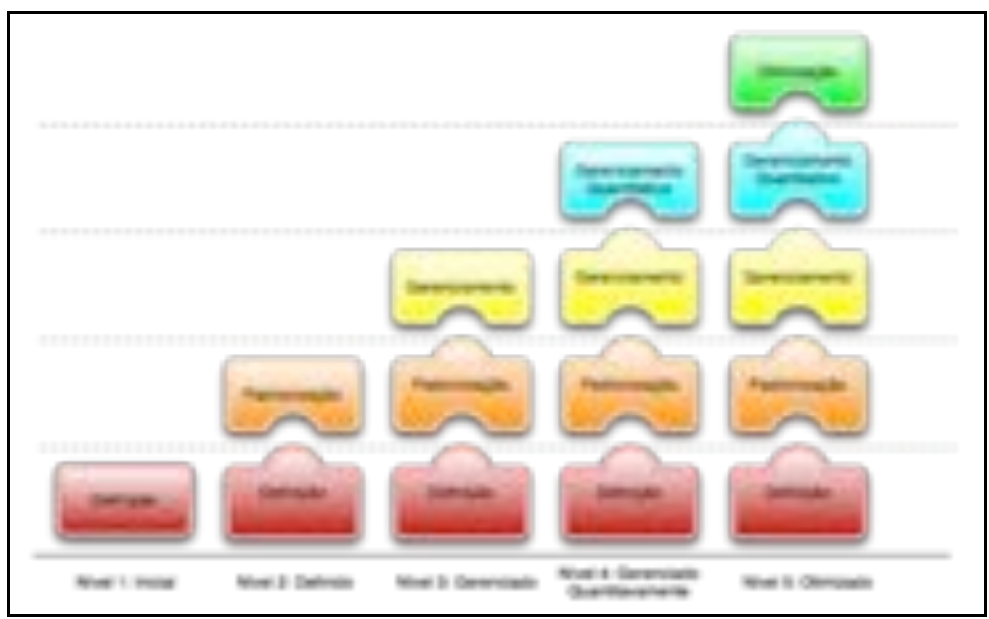

**Figura 20:** Acúmulo do grupo de processos por nível.

**Fonte:** ANTUNES, Ricardo Magno Acervo próprio.

A forma de avaliação proposta é a utilização de questionário utilizando a escala Likert em sua forma clássica. Cinco níveis de aceite de forma bipolar a determinada proposição. (FIG. 20)

**b. Proposição:** A organização possui padrões definidos e divulgados sobre técnicas de identificação de riscos.

- ⚪ Discordo completamente.

- ⚪ Discordo parcialmente.

- ⚪ Indiferente.

- ⚪ Concordo parcialmente.

- ⚪ Concordo completamente.

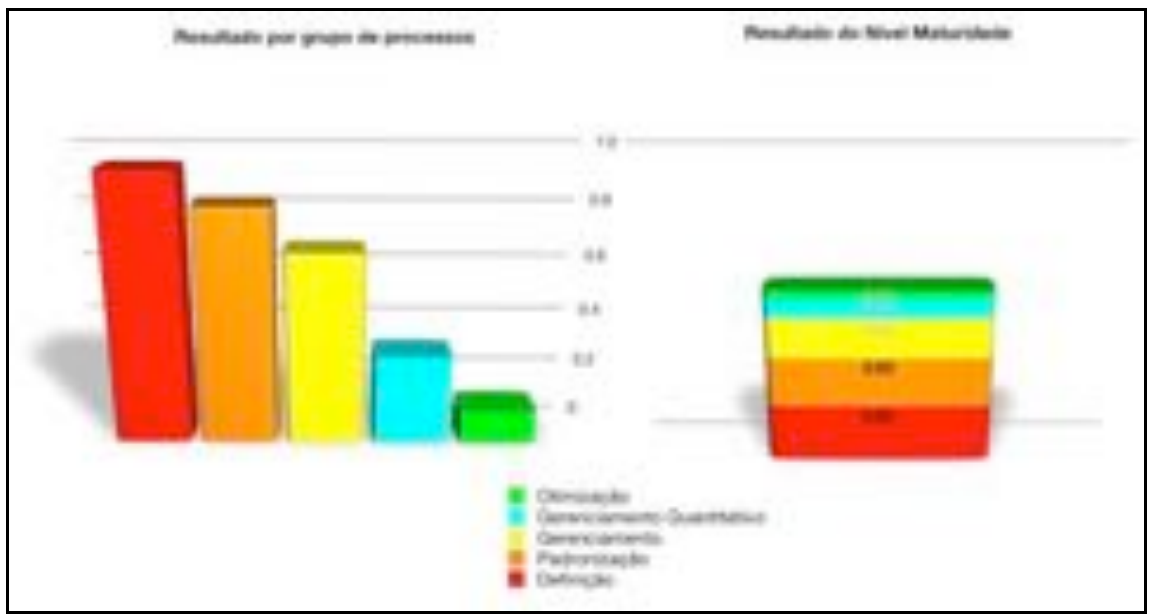

**Figura 20:** Forma de apresentação de resultados. Forma em que o modelo apresenta o resultado final ao avaliado.
**Fonte:** ANTUNES, Ricardo Magno Acervo próprio.

**c. Complexidade de aplicação:** a aplicação do modelo exige cuidados especiais como consultorias ou auditorias, estas podem ser internas ou externas.

Existe alguma certificação ou autorização necessária aos avaliadores por órgãos credenciados.

**d. Características dos níveis:**

• **Nível 1 – Inicial**

1. Não existe um processo formal de identificação de riscos, quando existente tem características individuais e não corporativas.

2. Baixa utilização de gerenciamento de risco.

3. Processo repetitivo e reativo aos riscos.

4. Trabalha-se com a incerteza.

5. Lições aprendidas não se propagam para novos projetos.

• **Nível 2 – Definido**

1. Políticas genéricas de risco e procedimentos formalizados e implementados.

2. Recursos definidos para gerenciamento de riscos.

3. Conscientização corporativa dos benefícios advindos do gerenciamento de riscos.

4. Pouco ou nenhum envolvimento da alta administração.

5. Capacitação básica da equipe de gerência de riscos.

6. Comunicação informal de riscos.

• **Nível 3 – Gerenciado**

1. Planejamento e gerenciamento de riscos baseado em lições aprendidas em projetos anteriores.

2. Área de gerenciamento de riscos definida e de conhecimento de toda a corporação.

3. Definição do gerente de riscos de cada projeto.

4. Comprometimento da alta administração.

5. Análise de risco qualitativa.

6. Gerenciamento de risco em todos os projetos.

7. Participação dos principais *stakeholders* no processo de gerenciamento de riscos.

8. Comunicação formal de riscos.

- **Nível 4 – Gerenciado Quantitativamente**

1. Quantificação dos riscos analisados.

2. Gerenciamento de oportunidades.

- **Nível 5 – Otimizado**

1. Uso ativo de informação sobre riscos para melhorar os processos organizacionais proporcionando vantagem competitiva.

2. Comprometimento *top-down* (de cima para baixo) na utilização do gerenciamento de riscos (dar exemplo).

3. Tomada de decisões baseada nas informações obtidas no gerenciamento de riscos.

4. Pro-atividade em gerenciamento de riscos encorajada e recompensada.

5. Reciclagens periódicas em gerenciamento de riscos.

6. Uso de ferramentas, técnicas e metodologias no estado da arte em gerenciamento de riscos.

7. Aprimoramento contínuo dos processos de gerenciamento de riscos.

### 3.3 Grupos de processos de gerenciamento de riscos por nível

### 3.3.1 Nível 1 – Inicial

### 3.3.1.1 Definição

A identificação de riscos determina quais os possíveis acontecimentos, seus efeitos nos objetivos do projeto e o modo que estes serão afetados. Os métodos e ferramentas utilizadas para a identificação dos riscos são próprios de cada indivíduo ou por demanda de cada projeto. Não existe metodologia consolidada para identificação de riscos.

A análise de riscos é superficial e baseada em informações restritas sem a utilização de histórico corporativo, mas em experiências dos envolvidos na analise. Frequentemente eventos especiais podem ocorrer sem identificação prévia. Por se tratarem de ações isoladas as lições aprendidas não são armazenas e difundidas a toda organização os riscos freqüentes enfrentados pela organização não são levados em conta.

A avaliação de riscos é o processo de comparação o risco identificado contra um critério definido a fim de determinas a magnitude do risco. Nem todos os interessados estão envolvidos na definição de critérios de avaliação de riscos o que comumente provoca desvios de superestimação ou subestimação na importância dos riscos.

### 3.3.2 Nível 2 – Definido – Definição +  Padronização

A identificação de riscos evolui e possui técnicas de identificação de riscos. Estas são padronizadas pela organização e os envolvidos estão treinados em seu uso. Os riscos identificados são conhecidos e entendidos pela organização, e suas potenciais conseqüências e suas possíveis variações são determinadas conforme a necessidade da decisão a ser tomada.

As técnicas de identificação de riscos incluem:

- *Brainstorming* (ou tempestade de idéias);
- *Checklists* (ou lista de verificação);
- Questionários circulados em uma gama de pessoas;
- Exame em projetos similares ou lições aprendidas;
- Entrevista com especialistas;
- Delphi.

Alguma documentação comum relativa a esse processo é produzida. Geralmente uma lista de possíveis riscos que podem afetar o sucesso do projeto (Processos posteriores determinarão as prioridades cujos quais serão tratados) Alguns itens presentes nessa documentação podem ser:

- Projeto: no projeto em que o risco esta presente.
- Elemento.
- Risco: o que e como.
- Gerente (dono do risco).
- Referência.
- Descrição e mecanismos.
- Premissas chaves.
- Fontes de informação.
- Lista de anexos.
- Revisor.
- Data.

### 3.3.2.1 Métodos

São utilizados para se determinar as conseqüências de cada risco, e estão presentes na análise e avaliação dos riscos. Também contemplam riscos secundários. Os riscos são priorizados conforme critérios estabelecidos pela organização.

### 3.3.2.2 Documentação gerada

A documentação gerada por este processo compreende uma lista de riscos detalhando o impacto no sucesso do projeto caso ocorram e a conseqüência de riscos secundários e uma lista priorizada de riscos assim como uma definição de níveis desta priorização.

### 3.3.2.3 Tratamento dos riscos

O objetivo do tratamento de riscos é determinar o que será feito em resposta ao risco identificado em ordem de reduzir a exposição geral ao risco. A entrada principal deste processo é a lista de risco priorizada, produzida no processo anterior.

### 3.3.2.4 Método

Identificar as opções para a redução das causas, conseqüências e/ou riscos secundários de cada risco considerado extremo, alto ou médio.

### 3.3.2.5 Documentação gerada

Plano de ação de riscos contendo as principais ações para cada risco considerado elevado assim como a nova classificação do risco.

### 3.3.2.6 Acompanhamento e revisão

Entrelace entre a gerência de riscos e outros processos de gerenciamento

### 3.3.2.7 Método

Implementar um processo de revisão com parte de regular da agenda de reuniões gerenciais.

### 3.3.2.8 Saídas

Revisões periódicas ao registro de riscos e lista de novas ações para o tratamento de riscos.

### 3.3.2.9 Documentação

Atualização do registro de risco como resultado do processo revisão.

### 3.3.3 Nível 3 – Gerenciado

A avaliação é realizada sobre riscos individuais relativo aos outros riscos e suportam o ajuste de prioridades destes riscos e também o alocamento de recursos para cada risco priorizado.

As estratégias para o tratamento de riscos são utilizadas de modo eficiente reduzindo o impacto de mais de um risco por ação.

### 3.3.3.1 Método

O método recomendado para o gerenciamento de riscos padronizado é consistente com os processos de gerenciamento de riscos definidos pela organização. A aplicação dos processos de gerenciamento d riscos é integrada e condizente com os processos e atividades do gerenciamento de projetos.

### 3.3.3.2 Avaliação qualitativa de riscos

As prioridades são utilizadas para determinar onde devem ser concentrados os maiores esforços para o tratamento de riscos identificados. Isto facilita a estruturação de planos e alocação de recursos.

### 3.3.3.3 Avaliação de riscos semi-quantitativo

Os processos descritos podem identificar potenciais sistemas, subsistemas, elementos ou estágios de riscos de nível considerados elevados de um projeto sem caracterizar explicitamente tais riscos assim como poderem ser utilizados para pesquisas de verificação de riscos em elementos no projeto.

### 3.3.3.4 Entradas

A informação utilizada na avaliação deste processo inclui informações vitais da documentação do projeto, como *Work Breakdown Structure* (WBS ou estrutura analítica do projeto, EAP) estratégia de execução de projetos, declaração de abertura, premissas de custo e prazo declaração de escopo, elementos técnicos e de engenharia, analises de viabilidade e econômicas e qualquer outra documentação relevante ao projeto ou a seu propósito. Outras informações como dados históricos, análises teóricas, análises e dados empíricos, opiniões especializadas e opinião de interessados também podem contribuir.

### 3.3.3.5 Método

- Desenvolver um sistema ou elemento estruturado apropriado para a exame do projeto.
- Uso de uma aproximação semi-quantitativa de avaliação de cada elemento do risco e suas consequências.

- Conversão de probabilidade e impacto do risco em uma prioridade inicial para o elemento.

### 3.3.3.6 Tratamento de riscos

A principal característica deste processo é o envolvimento com os planos de projeto e orçamento deste.

- **Método**

Identificar as opções para reduzir os impactos ou conseqüências de cada risco considerado elevado e determinar os benefícios e custos potencias das opções relacionadas.

1. Selecionar as melhores opções para o projeto.
2. Desenvolver e implementar um plano de ação de riscos detalhado.
3. Fazer provisões apropriadas de orçamento levando em conta os riscos e ações de tratamento.

### 3.3.3.7 Saídas

Plano de ação de riscos sumarizados para cara risco classificado como extremo ou alto conforme a escala da definida pela organização.

### 3.3.3.8 Acompanhamento e revisão

Acompanhamento e revisão contínua dos riscos garantindo que novos riscos sejam detectados e gerenciados, e planos de ação sejam implementados e progridam efetivamente.

- **Método**

Implementar um processo de revisão como parte de uma agenda regular de reuniões de gerenciamento. Realizar revisões detalhadas em marcos e fases significantes do projeto.

- **Saídas**

Revisões no registro de riscos incluindo uma lista de novos ações para tratamento de riscos.

- **Documentação**

Registro de riscos atualizado como resultado do processo de revisão.

- **Comunicação e relatórios**

Propósito: gerentes de projeto devem reportar o estado atual dos riscos e da gerência de riscos periodicamente de acordo a necessidade dos patrocinadores e interessados ou conforme políticas organizacionais.

- **Análise racional**

A alta gerência necessita possuir o entendimento da importância do gerenciamento de riscos. O objetivo principal dos relatórios de acompanhamento é fornecer este entendimento e a importância dos riscos frente aos objetivos do projeto.

- **Entradas**

O registro de riscos e planos de ação de suporte fornece a base para a maior parte do relatório de riscos do projeto.

- **Método**

Enviar relatórios regulares ou por demanda integrando-os ao plano de comunicação do projeto.

- **Saída**

Os relatórios fornecem sobre as ações de tratamento de riscos e indicações de tendências da incidência de riscos no projeto.

### 3.3.3.9 Planos e processos do projeto

- **Propósito**

O plano de gerenciamento de riscos do projeto especifica como a gerência de risco será conduzida no projeto, e o integra com outras atividades e processos do gerenciamento de projetos.

- **Análise Racional**

A gerência de riscos deve ser parte do negócio para todos envolvidos no projeto. O plano de gerenciamento de risco especifica como isto será alcançado pelo time de projeto.

- **Entradas**

O plano de gerenciamento de riscos é baseado em guias definidos pela organização, adaptadas as necessidades do projeto, e integrados ao plano de projeto e outros documentos.

- **Método**

Desenvolver o plano de gerenciamento de riscos em um estágio inicial do projeto, e mantê-lo atualizado a progressão do projeto durante em seu ciclo de vida.

- **Saídas**

Grandes projetos requerem um plano de gerenciamento de riscos mais detalhados, entretanto projetos considerados pequenos ou de risco baixo podem contemplar um plano mais simples. Dispensar a gerência de riscos pode ser considerado um grave erro na gerência de projetos, pois sendo um empreendimento único, isto é, nunca realizado antes, a existência de incertezas e a presença de riscos é inerente ao projeto.

### 3.3.3.10 Gerenciar oportunidades

- **Propósito**

A gerência de riscos foca-se nos possíveis desvios e mudanças do que é esperado ou foi planejado. Conforme a definição anterior de riscos, estes possíveis desvios podem ser positivos gerando conseqüências e ou resultados favoráveis ao projeto. Os riscos que podem favorecer o projeto são considerados oportunidades e podem ser identificadas e exploradas da mesma forma que os riscos de impacto negativo devem ser identificados e tratados.

- **Análise racional**

Se identificados e explorados a tempo, as oportunidades podem prover benefícios significantes ao projeto. Os benefícios podem ser identificados e processados utilizando uma extensão do processo de gerenciamento de riscos definido na organização ou integrando a este apenas realçando a bipolaridade do impacto (negativo para os riscos prejudiciais ou positivos para oportunidades) e incluindo as diferentes formas de tratamento das oportunidades; provocar, facilitar, aumentar o impacto.

- **Método**

Os processos gerais de gerenciamento de risco aplicam-se da mesma forma para o tratamento das oportunidades, requerendo apenas pequenos ajustes no impacto ao projeto e tratamentos ambos em sentidos contrários, isto é, o impacto passa a ser positivo ao invés de negativo, e os esforços são direcionados a provocar e não evitar o risco.

- **Saídas**

Os processos produzem um registro de oportunidades, análogo ao risco comum de riscos descrito anteriormente, e uma série de planos de ações paralelos aos anteriores.

- **Documentação**

A documentação gerada é a mesma que a produzida para os riscos comuns.

### 3.3.4 Nível 4 – Gerenciado quantitativamente (*Benchmarking*)

Os processos, por si mesmos, e as estratégias de tratamento de riscos são implementadas com foco na eficiência de custos.

### 3.3.4.1 Propósito

O modelo quantitativo de riscos em muitos componentes incertos possivelmente interage entre si, simultaneamente afetam a incerteza e o risco geral associado ao projeto. Isto lida com o propósito, desenvolvimento e análise dos modelos também com seu suporte a tomada de decisão.

### 3.3.4.2 Análise racional

A análise quantitativa de riscos propicia uma informação detalhada para a alta gerência na tomada de decisões sobre as incertezas do projeto. Ela contribui para decisões estratégicas sobre a aceitabilidade de opções do projeto, e a operacionalidade destas decisões em termos te alocação de recursos, metas, planos de contingência de modo consistente com o perfil da organização em relação a sua disposição em correr riscos.

### 3.3.4.3 Entradas

As entradas as modelos quantitativos incluem representações probabilistas de incerteza da ocorrência de eventos, distribuições de modelos de parâmetros, e correlações entre parâmetros.

### 3.3.4.4 Método

O modelo quantitativo de riscos envolve o estabelecimento do contexto e limites do modelo, estruturando as relações entre os riscos e o projeto, executando-o e validando-o em um processo interativo e interpretando suas saídas, utilizando-se de simulações.

### 3.3.4.5 Saídas

As saídas do modelo quantitativo de riscos incluem uma escala provável e realística de resultados esperáveis, o risco de exceder as metas especificadas em função de seus valores, a magnitude relativa das diferentes fontes de incerteza e os maiores causadores de risco para projeto.

### 3.3.4.6 Documentação

A documentação do modelo quantitativo de riscos deve gravar o processo pelo qual foi desenvolvido, sua estrutura, parâmetros e saídas, sua reconciliação com os envolvidos no projeto.

### 3.3.4.7 Tratamento de Risco

- **Método**

Realizar provisões adequadas a gerência de custos do projeto baseado no modelo quantitativo de riscos.

### 3.3.5 Nível 5 – Otimizado

### 3.3.5.1 Propósito

Facilitar um melhor gerenciamento dos riscos e melhoria contínua.

### 3.3.5.2 Método

Aplicação de ferramentas de qualidade e melhoria contínua em processos.

### 3.3.5.3 Entrada

Documentação gerada em projetos anteriores e padrões atuais dos processos de gerenciamento de riscos.

### 3.3.5.4 Saída

Padrões de processos atualizados conforme as lições aprendidas pela organização durante a execução de projetos anteriores.

### 3.4 Métrica para definição dos níveis

A pontuação não é cumulativa de um nível para outro. Ou seja, o cálculo é individual por nível.

A definição da pontuação mínima de cada nível de maturidade se baseia na fórmula matemática abaixo:

**Concordo Parcialmente x Nr. de questões do nível de maturidade = 75% cumprimento**

**OBS:** para manutenção do nível atual, é pré-requisito que seja mantida a pontuação dos níveis inferiores, já que neste modelo, o padrão adotado é reflexo do modelo por estágios. (FIG. 22)

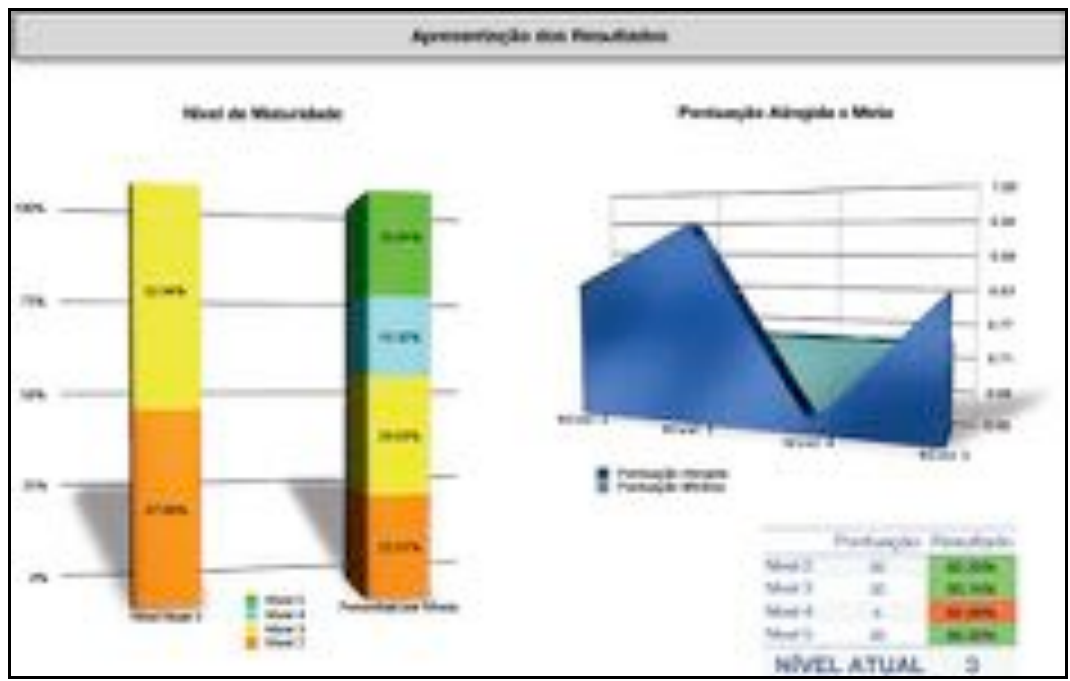

**Figura 22:** Apresentação dos resultados.
**Fonte:** BIRCHAL, Daniel Marssiére E ANTUNES, Ricardo Magno.  Acervo próprio.

# 4 ANÁLISE DOS RESULTADOS DA PESQUISA

## 4.1 Considerações

A metodologia da pesquisa utilizou abordagem quantitativa e qualitativa e o tipo de Pesquisa adotada foi a Bibliográfica (desenvolvida com base em publicações).

O universo são empresas que trabalham com projetos, de qualquer natureza e as amostras são formas submetidas a métricas de modelo matemático.

Elaboramos um questionário estruturado (quantitativo e qualitativo) para coleta de dados. Ele contém as perguntas necessárias para obter-se repostas aos objetivos da nossa pesquisa. As perguntas são simples, claras e objetivas, para que atinjam qualquer público das organizações. Os tipos de perguntas utilizadas foram:

- Fechadas;
- Com escalas de avaliação.

Os dados receberam tratamento matemático sobre a codificação /interpretação utilizada. Com base nos valores do questionário, foram calculados os níveis e percentuais destes, auxiliando os avaliadores no processo de análise e interpretação da maturidade.

# 5 CONCLUSÕES

A adoção de uma *"Gerência de Riscos em Projetos"* em uma organização não é um desafio simples e demanda longo período de tempo. Não deve ser considerado um mero processo de técnicas, treinamento de pessoas, e aquisição de software de controle. Ela deve ser utilizada nos mais simples ou complexos projetos.

O *Modelo de Maturidade em Gerência de Riscos em Projetos* aqui apresentado é um método estruturado para realçar os aspectos efetivos do Gerenciamento de Riscos. Permite às organizações medir sua gerência em riscos, face aos níveis de maturidade. Permite também identificar o que deve ser realizado a fim melhorar e aumentar sua habilidade de controlar o risco.

1. **Benefícios gerados:**

   a. **Para os praticantes:**

   - Constrói consenso e estabelece marcos;
   - *Benchmarking* das melhores práticas;
   - Comunicação clara para com a Diretoria, órgãos reguladores, agências de medição gerentes executivos, proprietários dos processos, funções/áreas de suporte.

   b. **Para os *Stakeholders*:**

   - Alinhamento do processo de Gerenciamento de Riscos na empresa;
   - Elimina esforço redundante e conecta as áreas de função /suporte com os proprietários dos processos;
   - Mede o valor do Gerenciamento de Riscos na empresa, baseado em suas prioridades, além de uma linguagem e visão compartilhada.

**c. Para as organizações:**

- Guia/direciona na melhoria contínua dos processos de *Gerência de Riscos*;

- Resolve processos de negócios ineficientes;

- Constrói/solidifica um processo repetitivo e escalável para melhores tomadas de decisão e reduz  custos (entender a causa raiz de um risco, é mais barato do que tratar o sintoma);

- Uma questão de aderência/compatibilidade pode levar a repensar a estratégia de negócio e encontrar uma oportunidade para gerar "receitas".

# 6 POSSÍVEIS DESDOBRAMENTOS

O uso do modelo permitirá também aos clientes, fornecedores e outras áreas da organização determinar como estão executando a *Gerência de Risco*, e pode ajudar no desenvolvimento de estratégias específicas para ir a um nível mais elevado da maturidade. Algum trabalho adicional é requerido para realçar os elementos diagnósticos do modelo, entretanto, a estrutura atual fornece uma ferramenta útil para as organizações interessadas numa aproximação formal à gerência de risco ou em melhorar sua aproximação existente.

# REFERÊNCIAS BIBLIOGRÁFICAS

**Sites consultados na Internet:**

- http://www.prmia.org/
- http://www.prince-officialsite.com/home/home.asp
- http://www.mor-officialsite.com/home/home.asp
- http://www.indg.com.br/projetos/downloads/texto sobre Modelo de Maturidade em Gerência de Projetos – MMGP
- http://www.simpros.com.br/simpros2002/Tutorial2
- http://www.cin.ufpe.br/~fabio/Gerenciamento
- http://www.ipma.ch
- http://www.ogc.gov.uk
- http://www.pmi.org
- http://opm3online.pmi.org
- http://www.logicmanager.com/contents/solutions/overview.php
- http://www.scielo.br/scielo.php
- http://www.snaptech.co.za/faq_root.html
- http://www.ilxsolutions.com

# GLOSSÁRIO

**ISO/IEC 15504** também conhecido como SPICE (Software Process Improvement and Capability Etermination) é uma "estrutura para análise do processo de desenvolvimento de software" pela Joint Technical Subcommittee (Subcomite Técnico Conjunto) entre ISO (International Organization for Standardization) e IEC (International Electrotechnical Commission). ISO/IEC 15504 deriva da ISO 12207 e utiliza várias das idéias do CMMI.

**Work Breakdown Structure** ou Estrutura Analítica do Projeto (EAP) é uma decomposição hierárquica orientada à entrega do trabalho a ser executado pela equipe do projeto, para atingir os objetivos do projeto e criar as entregas necessárias. A EAP organiza e define o escopo total do projeto. A EAP subdivide o trabalho do projeto em partes menores e mais facilmente gerenciáveis, em que cada nível descendente da EAP representa uma definição cada vez mais detalhada do trabalho do projeto. Fonte: Project Management Institute, Inc. Um Guia do Conjunto de Conhecimentos em Gerenciamento de Projetos Terceira edição (Guia PMBOK®)

***Risks Breakdown Structure*** ou Estrutura Analítica dos Riscos (EAR) é um diagrama que apresenta as categorias de riscos e suas subcategorias. É um elemento do plano de gerenciamento de riscos.

**Delphi, técnica ou método** O uso do Delphi foi recentemente envolvido no mundo dos negócios com a conexão a análise de riscos. Esta se preocupa com as incertezas associadas a novos projetos ou investimentos. Decisões, normalmente, são tomadas em ausência de informação adequada. O potencial de mercado de um novo produto é incerto e os custos de desenvolvimento podem exceder as estimativas. O pessoal de marketing em uma unidade operante da coorporação frequentemente apresenta um otimismo incandescente que negligencia a inteligência de competidores e a rápida capacidade de reação da concorrência. Engenheiros tendem a assumir custos de produtos complexos como uma função linear, isto é, uma somatória de custos dos componentes. Eles negligenciam as interações que resulta em um comportamento não linear: o custo total é muito maior que a soma das partes Assim, o custo é brutalmente subestimado. Um recente estudo de um grande número de projetos de desenvolvimento de produtos para defesa indica aproximadamente uma chance de 50% de estouro de custos. O método Delphi pode ser utilizado como uma vantagem para prover informações à análise de riscos. A parte mais crítica de tal análise é a subjetividade da distribuição probabilística assumida para as incertezas. Delphi poder servir para provar os pontos de vista do pessoal conectado com o projeto assim como os interessados (*stakeholders*). A anonimalidade é particularmente valiosa em uma alta estrutura de envolvimento onde indivíduos possam sentir-se constrangidos em expressar seus pontos de vista.

**Análise Quantitativa de Riscos:** Processo que determina as conseqüências dos riscos identificados para o projeto e sua probabilidade de ocorrência e dispõe os riscos por ordem de prioridade, de acordo com seu efeito sobre os objetivos do projeto.

**Análise Quantitativa de Riscos:** Processo que atribui probabilidades numéricas a cada risco identificado e examina seu impacto potencial aos objetivos do projeto.

**Árvores de decisão Diagramas:** que mostram a seqüência de decisões relacionadas e os resultados esperados na escolha de uma alternativa em detrimento de outra. É uma ferramenta de modelagem da Análise Quantitativa de Erros.

*Brainstorming:* Técnica de coleta de informações que é uma ferramenta e técnica do processo de Identificação de Riscos; consiste na reunião, num mesmo lugar, de especialistas em determinado assunto, membros da equipe, membros da equipe de gerenciamento de riscos e todos que possam vir a se beneficiar do processo para investigarem, juntos, possíveis eventos de risco.

**Categorias de riscos:** Identificam sistematicamente os riscos e servem de base para a sua compreensão. O uso de categorias de riscos ajuda a melhorar o processo de Identificação de Riscos, pois cria uma linguagem comum ou uma base para a descrição dos riscos a ser utilizada por todos. As categorias pertencem ao plano de gerenciamento de riscos.

**Controle e Monitoramento de Riscos:** Processo que envolve a resposta aos riscos à medida que ocorrem. Especifica como os riscos serão gerenciados, ao passo que o plano de respostas aos riscos descreve como serão implementadas as estratégias de resposta aos riscos caso um evento de risco se concretize. Implementa as respostas aos riscos de acordo com o plano.

**Escala cardinal:** Valores que podem ser lineares ou não-lineares e consultados no processo de Análise Qualitativa de Riscos.

**Escala de impacto:** Atribui um valor que descreva a gravidade do impacto de um possível risco usando um número cardinal ou valor numérico real.

**Escala ordinal:** Valores ordenados por critérios de classificação, como alto, médio e baixo. Citados no processo de Análise Qualitativa de Riscos.

**Gatilhos:** Sintomas que indicam que um evento de risco está prestes a ocorrer.

**Gerenciamento de Riscos do Projeto:** Uma das nove áreas de conhecimento do gerenciamento de projetos. Tem como objetivo a identificação e planejamento de riscos potenciais que possam afetar o projeto e compreende os seguintes processos: Planejamento do Gerenciamento dos Riscos, Identificação de Riscos, Análises Qualitativa e Quantitativa de Riscos, Planejamento de Resposta aos Riscos e Controle e Monitoramento de Riscos.

**Identificação de Riscos:** Processo que identifica os potenciais riscos do projeto e documenta suas características.

**Impacto:** A quantidade de danos ou oportunidades que um evento de risco oferece ao projeto.

**Probabilidade:** Chances de que determinado evento de risco ocorra.

**Intensificação:** Estratégia de resposta empregada quando o risco representa uma oportunidade para o projeto.

**Participação:** Estratégia do Plano de Respostas aos Riscos empregada quando o risco representa uma oportunidade para o projeto.

**Matriz de probabilidade e impacto:** Define a combinação de probabilidade e impacto que ajuda a determinar quais riscos precisam de planos de resposta detalhados. Ela é definida durante o processo de Planejamento do Gerenciamento de Riscos e incluída no plano de gerenciamento de riscos. A matriz é utilizada no processo de Análise Qualitativa de Riscos para atribuir uma classificação geral a cada um dos riscos identificados no projeto.

**Medida preventiva:** Envolve qualquer coisa que reduza os possíveis impactos dos eventos de risco. Os planos de contingências e as respostas aos riscos são exemplos de medidas preventivas.

**Planejamento de Respostas aos Riscos:** Processo que define as medidas a tomar para reduzir ameaças e tirar vantagens das oportunidades.

**Planejamento de contingências:** Estratégia de respostas a riscos que envolvem o planejamento de alternativas para lidar com os riscos, caso se concretizem.

**Planejamento de Gerenciamento de Riscos:** Processo que determina como os riscos serão gerenciados no projeto.

**Plano de gerenciamento de riscos:** Esmiúça como os processo de gerenciamento de riscos serão definidos, monitorados e controlados no decorrer da vida do projeto.

**Registro de riscos:** Saída do processo de Identificação dos Riscos que contém uma lista dos riscos identificados. Quando atualizado, as novas informações se tornam saída do que ainda resta dos processos de Riscos.

**Risco residual:** Risco que permanece após a implementação de uma estratégia de resposta a riscos.

**Riscos secundários:** Riscos resultantes da implementação de uma resposta a riscos.

**Solução alternativa:** Resposta não-planejada a um evento de risco que era desconhecido ou não fora identificado, ou resposta não-planejada a um risco aceito anteriormente.

**Tolerância a riscos:** Nível no qual os *stakeholders* se sentem confortáveis para encarar um risco porque os benefícios a serem alcançados superam o que poderia ser perdido.

**Transferência:** Ferramenta e técnica do processo de Planejamento de Respostas a Riscos que transfere as conseqüências de determinado risco para um terceiro.

# APÊNDICES

**APÊNDICE 1: QUESTIONÁRIO**

**Pontuação para as opções:**

Discordo Completamente (0)

Discordo Parcialmente (1)

Indiferente (2)

Concordo Parcialmente (3)

Concordo Completamente (4)

**Nível 2 – Definido**

1. O assunto *"Gerenciamento de Riscos"* é aceito por parte dos gerentes de Projetos.

Discordo Completamente

Discordo Parcialmente

Indiferente

Concordo Parcialmente

Concordo Completamente

2. Existe um processo formal de identificação de riscos.

Discordo Completamente

Discordo Parcialmente

Indiferente

Concordo Parcialmente

Concordo Completamente

3. A reação aos riscos é pro ativa.

Discordo Completamente

Discordo Parcialmente

Indiferente

Concordo Parcialmente

Concordo Completamente

4. Existem políticas genéricas de risco e procedimentos formalizados e implementados.

Discordo Completamente

Discordo Parcialmente

Indiferente

Concordo Parcialmente

Concordo Completamente

5. Existem recursos alocados para gerenciar de riscos.

Discordo Completamente

Discordo Parcialmente

Indiferente

Concordo Parcialmente

Concordo Completamente

6. Há conscientização corporativa dos benefícios provenientes do gerenciamento de riscos.

Discordo Completamente

Discordo Parcialmente

Indiferente

Concordo Parcialmente

Concordo Completamente

7. A alta administração da empresa está envolvida.

Discordo Completamente

Discordo Parcialmente

Indiferente

Concordo Parcialmente

Concordo Completamente

8. A equipe foi capacitada para gerenciamento de riscos.

Discordo Completamente

Discordo Parcialmente

Indiferente

Concordo Parcialmente

Concordo Completamente

9.  O canal de comunicação para informações referentes a riscos é informal.

Discordo Completamente

Discordo Parcialmente

Indiferente

Concordo Parcialmente

Concordo Completamente

## Nível 3 – Gerenciado

1.  O planejamento e gerenciamento de riscos baseado em lições aprendidas em projetos anteriores

Discordo Completamente

Discordo Parcialmente

Indiferente

Concordo Parcialmente

Concordo Completamente

2.  Existe uma área de gerenciamento de riscos definida e de conhecimento de toda a corporação.

Discordo Completamente

Discordo Parcialmente

Indiferente

Concordo Parcialmente

Concordo Completamente

3.  Para cada projeto é definido um gerente de riscos

Discordo Completamente

Discordo Parcialmente

Indiferente

Concordo Parcialmente

Concordo Completamente

4.  A alta administração está comprometida, apoiando e encorajando o gerenciamento de riscos.

Discordo Completamente

Discordo Parcialmente

Indiferente

Concordo Parcialmente

Concordo Completamente

5. A análise de riscos é qualitativa.

Discordo Completamente

Discordo Parcialmente

Indiferente

Concordo Parcialmente

Concordo Completamente

6. Gerenciamento de risco é feito em todos os projetos.

Discordo Completamente

Discordo Parcialmente

Indiferente

Concordo Parcialmente

Concordo Completamente

7. Os principais fornecedores são envolvidos no processo de gerenciamento de riscos.

Discordo Completamente

Discordo Parcialmente

Indiferente

Concordo Parcialmente

Concordo Completamente

8. Existe um orçamento anual corporativo definido para gerenciamento de riscos.

Discordo Completamente

Discordo Parcialmente

Indiferente

Concordo Parcialmente

Concordo Completamente

9. Desenvolver um plano de ação para riscos

Discordo Completamente

Discordo Parcialmente

Indiferente

Concordo Parcialmente

Concordo Completamente

## Nível 4 – Gerenciado quantitativo

1.  A análise de riscos é quantitativa.

Discordo Completamente

Discordo Parcialmente

Indiferente

Concordo Parcialmente

Concordo Completamente

2. Em relação ao assunto "Gerenciamento de Riscos" por parte dos gerentes de Projeto oportunidades e ameaças são identificadas no gerenciamento de riscos.

Discordo Completamente

Discordo Parcialmente

Indiferente

Concordo Parcialmente

Concordo Completamente

3. Identificar os benefícios e custos para cada resposta aos riscos:

Discordo Completamente

Discordo Parcialmente

Indiferente

Concordo Parcialmente

Concordo Completamente

4.  Utilizar ferramenta de medição de área de inferência global do risco.

Discordo Completamente

Discordo Parcialmente

Indiferente

Concordo Parcialmente

Concordo Completamente

**Nível 5 – Otimizado**

1. A cultura de atenção ao risco é pró-ativa.

Discordo Completamente

Discordo Parcialmente

Indiferente

Concordo Parcialmente

Concordo Completamente

2. Informação sobre riscos são ativamente utilizadas para melhorar os processos organizacionais proporcionando vantagem competitiva.

Discordo Completamente

Discordo Parcialmente

Indiferente

Concordo Parcialmente

Concordo Completamente

3. Há comprometimento *top-down* na utilização do gerenciamento de riscos.

Discordo Completamente

Discordo Parcialmente

Indiferente

Concordo Parcialmente

Concordo Completamente

4. As informações obtidas no gerenciamento de riscos influenciam na tomada de decisões.

Discordo Completamente

Discordo Parcialmente

Indiferente

Concordo Parcialmente

Concordo Completamente

5. Pro-atividade em gerenciamento de riscos encorajada e recompensada pela alta administração.

Discordo Completamente

Discordo Parcialmente

Indiferente

Concordo Parcialmente

Concordo Completamente

6. Reciclagens periódicas em gerenciamento de riscos.

Discordo Completamente

Discordo Parcialmente

Indiferente

Concordo Parcialmente

Concordo Completamente

7. São utilizadas ferramentas, técnicas e metodologias no estado da arte em gerenciamento de riscos.

Discordo Completamente

Discordo Parcialmente

Indiferente

Concordo Parcialmente

Concordo Completamente

8. Os processos de gerenciamento de riscos são continuamente revistos e melhorados.

Discordo Completamente

Discordo Parcialmente

Indiferente

Concordo Parcialmente

Concordo Completamente

9. Em relação à visibilidade da organização no mercado, é tida como *benchmark* em gerenciamento de riscos.

Discordo Completamente

Discordo Parcialmente

Indiferente

Concordo Parcialmente

Concordo Completamente

10. São utilizadas ferramentas de Gestão de Qualidade, como apoio ao Gerenciamento de Riscos.

Discordo Completamente

Discordo Parcialmente

Indiferente

Concordo Parcialmente

Concordo Completamente

**APÊNDICE 2: Questionário de avaliação de maturidade em gerenciamento de projetos.**

**Protocolo de Questionário**

Baseado no Modelo PMMM: Nível 2 de Maturidade – Fases do Ciclo de Vida -

Perfil do Entrevistado:

Nome:

Cargo / Função Atual:

Tempo na Função Atual:

Tempo na Empresa:

**Sessão I - Avaliação da maturidade em gestão de projetos**

Você encontrará 20 questões que o ajudarão a definir o grau de maturidade que, segundo a sua avaliação e percepção, sua empresa atingiu. Abaixo de cada questão, você deverá assinalar o número correspondente a sua avaliação / percepção, segundo a legenda exibida a seguir:

**(-3): Discordo Totalmente**

**(-2): Discordo**

**(-1): Discordo Parcialmente**

**( 0 ): Sem opinião**

**(+1): Concordo Parcialmente**

**(+2): Concordo**

**(+3): Concordo Totalmente**

A pontuação para cada uma das questões varia de (-3) a (+3) e será posteriormente utilizada para a avaliação dos resultados. Dessa forma, solicita-se que o entrevistado marque com um "X" a resposta para cada uma das 20 questões apresentadas a seguir. Seja, por favor, o mais honesto possível nas suas respostas. Marque a resposta que você considera correta, não aquela que você desejaria ou imaginaria que fosse a mais adequada.

**Questão 1:**

Minha empresa reconhece a necessidade da gestão de projetos. Esta necessidade é reconhecida em todos os níveis da gerência, inclusive pela gerência sênior.

**Discordo Totalmente (-3) (-2) (-1) (0) (+1) (+2) (+3) Concordo Totalmente**

**Questão 2:**

Minha empresa tem um sistema para gerenciar tanto o custo quanto o cronograma dos projetos. O sistema requer números de encargos financeiros e códigos de conta contábil. O sistema informa variações em relação aos objetivos planejados.

**Discordo Totalmente (-3) (-2) (-1) (0) (+1) (+2) (+3) Concordo Totalmente**

**Questão 3:**

Minha empresa tem reconhecido as vantagens possíveis de serem alcançadas através da implementação da gestão de projetos. Estes benefícios são reconhecidos em todos os níveis gerenciais, incluindo a gerência sênior.

**Discordo Totalmente (-3) (-2) (-1) (0) (+1) (+2) (+3) Concordo Totalmente**

**Questão 4:**

Minha empresa ou departamento tem uma metodologia facilmente identificável de gestão de projetos que utiliza o conceito de fases ou ciclo de vida de um projeto.

**Discordo Totalmente (-3) (-2) (-1) (0) (+1) (+2) (+3) Concordo Totalmente**

**Questão 5:**

Nossos executivos apóiam ostensivamente a gestão de projetos por meio de palestras, curso, artigos e inclusive pela presença ocasional em reuniões e relatórios da equipe de projetos.

**Discordo Totalmente (-3) (-2) (-1) (0) (+1) (+2) (+3) Concordo Totalmente**

**Questão 6:**

Minha empresa tem o compromisso com o planejamento antecipado visando à qualidade. Tentamos fazer sempre o melhor possível em matéria de planejamento.
**Discordo Totalmente (-3) (-2) (-1) (0) (+1) (+2) (+3) Concordo Totalmente**

**Questão 7:**

Nossos gerentes de área de níveis médio e inicial apóiam por completo e de forma ostensiva o processo de gestão de projetos.

**Discordo Totalmente (-3) (-2) (-1) (0) (+1) (+2) (+3) Concordo Totalmente**

**Questão 8:**

Minha empresa faz o possível para minimizar os desvios de escopo (por exemplo, mudança de escopo ou redefinição da extensão do escopo) em nossos projetos.
**Discordo Totalmente (-3) (-2) (-1) (0) (+1) (+2) (+3) Concordo Totalmente**

**Questão 9:**

Nossos gerentes de área estão comprometidos não apenas com a gestão dos projetos, mas também com o cumprimento dos prazos estabelecidos para a conclusão dos objetivos.

**Discordo Totalmente (-3) (-2) (-1) (0) (+1) (+2) (+3) Concordo Totalmente**

**Questão 10:**

Os executivos em minha empresa têm bom conhecimento dos princípios de gestão de projetos.
**Discordo Totalmente (-3) (-2) (-1) (0) (+1) (+2) (+3) Concordo Totalmente**

**Questão 11:**

Minha empresa selecionou um ou mais softwares para serem utilizados como sistema de controle dos projetos.

**Discordo Totalmente (-3) (-2) (-1) (0) (+1) (+2) (+3) Concordo Totalmente**

**Questão 12:**

Nossos gerentes de área de níveis médio e inicial foram treinados e instruídos em gestão de projetos.

**Discordo Totalmente (-3) (-2) (-1) (0) (+1) (+2) (+3) Concordo Totalmente**

**Questão 13:**

Nossos executivos compreendem o conceito de responsabilidade e atuam como patrocinadores *("sponsors")* em determinados projetos.

**Discordo Totalmente (-3) (-2) (-1) (0) (+1) (+2) (+3) Concordo Totalmente**

**Questão 14:**

Nossos executivos reconheceram ou identificaram as aplicações da gestão de projetos nas várias divisões (demais unidades de *Lighting*) do nosso empreendimento.

**Discordo Totalmente (-3) (-2) (-1) (0) (+1) (+2) (+3) Concordo Totalmente**

**Questão 15:**

Minha empresa conseguiu integrar com sucesso o controle de custo e cronogramas tanto para a gestão de projetos quanto para relatórios de *follow-up.*

**Discordo Totalmente (-3) (-2) (-1) (0) (+1) (+2) (+3) Concordo Totalmente**

**Questão 16:**

Minha empresa desenvolveu um currículo de gestão de projetos (por exemplo, mais do que um ou dois cursos de capacitação) para o aperfeiçoamento das qualificações de nossos colaboradores em gestão de projetos.

**Discordo Totalmente (-3) (-2) (-1) (0) (+1) (+2) (+3) Concordo Totalmente**

**Questão 17:**

Nossos executivos reconheceram o que precisa ser feito a fim de ser alcançada a maturidade em gestão de projetos.

**Discordo Totalmente (-3) (-2) (-1) (0) (+1) (+2) (+3) Concordo Totalmente**

**Questão 18:**

Minha empresa considera e trata a gestão de projetos como profissão, e não apenas como tarefa de tempo parcial ou, quando requerido, tempo integral.

**Discordo Totalmente (-3) (-2) (-1) (0) (+1) (+2) (+3) Concordo Totalmente**

**Questão 19:**

Nossos gerentes de área e nível médio estão dispostos a liberar seus funcionários para o treinamento em gestão de projetos.

**Discordo Totalmente (-3) (-2) (-1) (0) (+1) (+2) (+3) Concordo Totalmente**

**Questão 20:**

Nossos executivos têm demonstrado disposição para mudança na maneira tradicional de conduzir negócios para chegar à maturidade em gestão de projetos.

**Discordo Totalmente (-3) (-2) (-1) (0) (+1) (+2) (+3) Concordo Totalmente**

# AVALIAÇÃO/MÉTRICA DA MATURIDADE EM GESTÃO DE PROJETOS

## FORMULÁRIO PARA APURAÇÃO DA PONTUAÇÃO POR FASE DO CICLO DE VIDA

Para cada uma das questões (questões de número 1 até 20) da sessão I, você assinalou uma resposta com seu respectivo valor de pontuação, variando de (-3) até (+3). Nos espaços apropriados, indicados nas tabelas a seguir, favor transcrever o valor assinalado ao lado do número corresponde às questões respondidas.

Fase: Embrionária

| Número da Questão | Valor da Pontuação |
|---|---|
| # 1 | _______ |
| # 3 | _______ |
| # 14 | _______ |
| # 17 | _______ |
| Total | _______ |

Fase: Aceitação – Alta Direção

| Número da Questão | Valor da Pontuação |
|---|---|
| # 5 | _______ |
| # 10 | _______ |
| # 13 | _______ |
| # 20 | _______ |
| Total | _______ |

Fase: Aceitação - Gerência

| Número da Questão | Valor da Pontuação |
|---|---|
| # 7 | _______ |
| # 9 | _______ |
| # 12 | _______ |
| # 19 | _______ |
| Total | _______ |

## Fase: Crescimento

| Número da Questão | Valor da Pontuação |
|---|---|
| # 4 | _______ |
| # 6 | _______ |
| # 8 | _______ |
| # 11 | _______ |
| Total | _______ |

## Fase: Maturidade

| Número da Questão | Valor da Pontuação |
|---|---|
| # 2 | _______ |
| # 15 | _______ |
| # 16 | _______ |
| # 18 | _______ |
| Total | _______ |

# ANEXOS

**ANEXO 1: CMMI for Development - Risk Management a Project Management Process Area at Maturity Level 3**

# RISK MANAGEMENT

A Project Management Process Area at Maturity Level 3

## Purpose

The purpose of Risk Management (RSKM) is to identify potential problems before they occur so that risk-handling activities can be planned and invoked as needed across the life of the product or project to mitigate adverse impacts on achieving objectives.

## Introductory Notes

Risk management is a continuous, forward-looking process that is an important part of management. Risk management should address issues that could endanger achievement of critical objectives. A continuous risk management approach is applied to effectively anticipate and mitigate the risks that may have a critical impact on the project.

Effective risk management includes early and aggressive risk identification through the collaboration and involvement of relevant stakeholders, as described in the stakeholder involvement plan addressed in the Project Planning process area. Strong leadership across all relevant stakeholders is needed to establish an environment for the free and open disclosure and discussion of risk.

Risk management must consider both internal and external sources for cost, schedule, and performance risk as well as other risks. Early and aggressive detection of risk is important because it is typically easier, less costly, and less disruptive to make changes and correct work efforts during the earlier, rather than the later, phases of the project.

Risk management can be divided into three parts: defining a risk management strategy; identifying and analyzing risks; and handling identified risks, including the implementation of risk mitigation plans when needed.

As represented in the Project Planning and Project Monitoring and Control process areas, organizations may initially focus simply on risk identification for awareness, and react to the realization of these risks as they occur. The Risk Management process area describes an evolution of these specific practices to systematically plan, anticipate, and mitigate risks to proactively minimize their impact on the project.

Although the primary emphasis of the Risk Management process area is on the project, the concepts can also be applied to manage organizational risks.

## Related Process Areas

*Refer to the Project Planning process area for more information about identification of project risks and planning for involvement of relevant stakeholders.*

*Refer to the Project Monitoring and Control process area for more information about monitoring project risks.*

*Refer to the Decision Analysis and Resolution process area for more information about using a formal evaluation process to evaluate alternatives for selection and mitigation of identified risks.*

### Specific Goal and Practice Summary

SG 1 Prepare for Risk Management
    SP 1.1    Determine Risk Sources and Categories
    SP 1.2    Define Risk Parameters
    SP 1.3    Establish a Risk Management Strategy
SG 2 Identify and Analyze Risks
    SP 2.1    Identify Risks
    SP 2.2    Evaluate, Categorize, and Prioritize Risks
SG 3 Mitigate Risks
    SP 3.1    Develop Risk Mitigation Plans
    SP 3.2    Implement Risk Mitigation Plans

## Specific Practices by Goal

### SG 1    Prepare for Risk Management

*Preparation for risk management is conducted.*

Preparation is conducted by establishing and maintaining a strategy for identifying, analyzing, and mitigating risks. This is typically documented in a risk management plan. The risk management strategy addresses the specific actions and management approach used to apply and control the risk management program. This includes identifying the sources of risk; the scheme used to categorize risks; and the parameters used to evaluate, bound, and control risks for effective handling.

### SP 1.1    Determine Risk Sources and Categories

*Determine risk sources and categories.*

Identification of risk sources provides a basis for systematically examining changing situations over time to uncover circumstances that impact the ability of the project to meet its objectives. Risk sources are both internal and external to the project. As the project progresses, additional sources of risk may be identified. Establishing categories for risks provides a mechanism for collecting and organizing risks as well as ensuring appropriate scrutiny and management attention for those risks that can have more serious consequences on meeting project objectives.

**Typical Work Products**

1.  Risk source lists (external and internal)

2.  Risk categories list

**Subpractices**

1.  Determine risk sources.

    Risk sources are the fundamental drivers that cause risks within a project or organization. There are many sources of risks, both internal and external, to a project. Risk sources identify common areas where risks may originate. Typical internal and external risk sources include the following:

    - Uncertain requirements

    - Unprecedented efforts—estimates unavailable

    - Infeasible design

    - Unavailable technology

    - Unrealistic schedule estimates or allocation

    - Inadequate staffing and skills

    - Cost or funding issues

    - Uncertain or inadequate subcontractor capability

    - Uncertain or inadequate vendor capability

    - Inadequate communication with actual or potential customers or with their representatives

    - Disruptions to continuity of operations

    Many of these sources of risk are often accepted without adequate planning. Early identification of both internal and external sources of risk can lead to early identification of risks. Risk mitigation plans can then be implemented early in the project to preclude occurrence of the risks or reduce the consequences of their occurrence.

2.  Determine risk categories.

    Risk categories reflect the "bins" for collecting and organizing risks. A reason for identifying risk categories is to help in the future consolidation of the activities in the risk mitigation plans.

    > The following factors may be considered when determining risk categories:
    >
    > - The phases of the project's lifecycle model (e.g., requirements, design, manufacturing, test and evaluation, delivery, and disposal)
    >
    > - The types of processes used
    >
    > - The types of products used
    >
    > - Program management risks (e.g., contract risks, budget/cost risks, schedule risks, resources risks, performance risks, and supportability risks)

    A risk taxonomy can be used to provide a framework for determining risk sources and categories.

**SP 1.2      Define Risk Parameters**

*Define the parameters used to analyze and categorize risks, and the parameters used to control the risk management effort.*

Parameters for evaluating, categorizing, and prioritizing risks include the following:

• Risk likelihood (i.e., probability of risk occurrence)

• Risk consequence (i.e., impact and severity of risk occurrence)

• Thresholds to trigger management activities

Risk parameters are used to provide common and consistent criteria for comparing the various risks to be managed. Without these parameters, it would be very difficult to gauge the severity of the unwanted change caused by the risk and to prioritize the necessary actions required for risk mitigation planning.

**Typical Work Products**

1. Risk evaluation, categorization, and prioritization criteria

2. Risk management requirements (e.g., control and approval levels, and reassessment intervals)

**Subpractices**

1. Define consistent criteria for evaluating and quantifying risk likelihood and severity levels.

   Consistently used criteria (e.g., the bounds on the likelihood and severity levels) allow the impacts of different risks to be commonly understood, to receive the appropriate level of scrutiny, and to obtain the management attention warranted. In managing dissimilar risks (e.g., personnel safety versus environmental pollution), it is important to ensure consistency in end result (e.g., a high risk of environmental pollution is as important as a high risk to personnel safety).

2. Define thresholds for each risk category.

   For each risk category, thresholds can be established to determine acceptability or unacceptability of risks, prioritization of risks, or triggers for management action.

   > Examples of thresholds include the following:
   >
   > • Project-wide thresholds could be established to involve senior management when product costs exceed 10 percent of the target cost or when Cost Performance Indexes (CPIs) fall below 0.95.
   >
   > • Schedule thresholds could be established to involve senior management when Schedule Performance Indexes (SPIs) fall below 0.95.
   >
   > • Performance thresholds could be set to involve senior management when specified key items (e.g., processor utilization or average response times) exceed 125 percent of the intended design.

   These may be refined later, for each identified risk, to establish points at which more aggressive risk monitoring is employed or to signal the implementation of risk mitigation plans.

3. Define bounds on the extent to which thresholds are applied against or within a category.

There are few limits to which risks can be assessed in either a quantitative or qualitative fashion. Definition of bounds (or boundary conditions) can be used to help scope the extent of the risk management effort and avoid excessive resource expenditures. Bounds may include exclusion of a risk source from a category. These bounds can also exclude any condition that occurs less than a given frequency.

**SP 1.3    Establish a Risk Management Strategy**

*Establish and maintain the strategy to be used for risk management.*

A comprehensive risk management strategy addresses items such as the following:

- The scope of the risk management effort
- Methods and tools to be used for risk identification, risk analysis, risk mitigation, risk monitoring, and communication
- Project-specific sources of risks
- How these risks are to be organized, categorized, compared, and consolidated
- Parameters, including likelihood, consequence, and thresholds, for taking action on identified risks
- Risk mitigation techniques to be used, such as prototyping, piloting, simulation, alternative designs, or evolutionary development
- Definition of risk measures to monitor the status of the risks
- Time intervals for risk monitoring or reassessment

The risk management strategy should be guided by a common vision of success that describes the desired future project outcomes in terms of the product that is delivered, its cost, and its fitness for the task. The risk management strategy is often documented in an organizational or a project risk management plan. The risk management strategy is reviewed with relevant stakeholders to promote commitment and understanding.

**Typical Work Products**

1. Project risk management strategy

**SG 2    Identify and Analyze Risks**

*Risks are identified and analyzed to determine their relative importance.*

The degree of risk impacts the resources assigned to handle an identified risk and the determination of when appropriate management attention is required.

Analyzing risks entails identifying risks from the internal and external sources identified and then evaluating each identified risk to determine its likelihood and consequences. Categorization of the risk, based on an evaluation against the established risk categories and criteria developed for the risk management strategy, provides the information needed for risk handling. Related risks may be grouped for efficient handling and effective use of risk management resources.

**SP 2.1    Identify Risks**

*Identify and document the risks.*

---

**IPPD Addition**

The particular risks associated with conducting the project using integrated teams should be considered, such as risks associated with loss of inter-team or intra-team coordination.

---

The identification of potential issues, hazards, threats, and vulnerabilities that could negatively affect work efforts or plans is the basis for sound and successful risk management. Risks must be identified and described in an understandable way before they can be analyzed and managed properly. Risks are documented in a concise statement that includes the context, conditions, and consequences of risk occurrence.

Risk identification should be an organized, thorough approach to seek out probable or realistic risks in achieving objectives. To be effective, risk identification should not be an attempt to address every possible event regardless of how highly improbable it may be. Use of the categories and parameters developed in the risk management strategy, along with the identified sources of risk, can provide the discipline and streamlining appropriate to risk identification. The identified risks form a baseline to initiate risk management activities. The list of risks should be reviewed periodically to reexamine possible sources of risk and changing conditions to uncover sources and risks previously overlooked or nonexistent when the risk management strategy was last updated.

Risk identification activities focus on the identification of risks, not placement of blame. The results of risk identification activities are not used by management to evaluate the performance of individuals.

There are many methods for identifying risks. Typical identification methods include the following:

- Examine each element of the project work breakdown structure to uncover risks.

- Conduct a risk assessment using a risk taxonomy.

- Interview subject matter experts.

- Review risk management efforts from similar products.

- Examine lessons-learned documents or databases.

- Examine design specifications and agreement requirements.

**Typical Work Products**

1. List of identified risks, including the context, conditions, and consequences of risk occurrence

**Subpractices**

1. Identify the risks associated with cost, schedule, and performance.

   Cost, schedule, and performance risks should be examined to the extent that they impact project objectives. There may be potential risks discovered that are outside the scope of the project's objectives but vital to customer interests. For example, the risks in development costs, product acquisition costs, cost of spare (or replacement) products, and product disposition (or disposal) costs have design implications. The customer may not have considered the full cost of supporting a fielded product or using a delivered service. The customer should be informed of such risks, but actively managing those risks may not be necessary. The mechanisms for making such decisions should be examined at project and organization levels and put in place if deemed appropriate, especially for risks that impact the ability to verify and validate the product.

   In addition to the cost risks identified above, other cost risks may include those associated with funding levels, funding estimates, and distributed budgets.

   Schedule risks may include risks associated with planned activities, key events, and milestones.

Performance risks may include risks associated with the following:

- Requirements
- Analysis and design
- Application of new technology
- Physical size
- Shape
- Weight
- Manufacturing and fabrication
- Functional performance and operation
- Verification
- Validation
- Performance maintenance attributes

Performance maintenance attributes are those characteristics that enable an in-use product or service to provide originally required performance, such as maintaining safety and security performance.

There are other risks that do not fall into cost, schedule, or performance categories.

Examples of these other risks include the following:

- Risks associated with strikes
- Diminishing sources of supply
- Technology cycle time
- Competition

2. Review environmental elements that may impact the project.

Risks to a project that frequently are missed include those supposedly outside the scope of the project (i.e., the project does not control whether they occur but can mitigate their impact), such as weather, natural or manmade disasters that affect continuity of operations, political changes, and telecommunications failures.

3. Review all elements of the work breakdown structure as part of identifying risks to help ensure that all aspects of the work effort have been considered.

4. Review all elements of the project plan as part of identifying risks to help ensure that all aspects of the project have been considered.

*Refer to the Project Planning process area for more information about identifying project risks.*

5. Document the context, conditions, and potential consequences of the risk.

Risks statements are typically documented in a standard format that contains the risk context, conditions, and consequences of occurrence. The risk context provides additional information such that the intent of the risk can be easily understood. In documenting the context of the risk, consider the relative time frame of the risk, the circumstances or conditions surrounding the risk that has brought about the concern, and any doubt or uncertainty.

6. Identify the relevant stakeholders associated with each risk.

## SP 2.2     Evaluate, Categorize, and Prioritize Risks

***Evaluate and categorize each identified risk using the defined risk categories and parameters, and determine its relative priority.***

The evaluation of risks is needed to assign relative importance to each identified risk, and is used in determining when appropriate management attention is required. Often it is useful to aggregate risks based on their interrelationships, and develop options at an aggregate level. When an aggregate risk is formed by a roll up of lower level risks, care must be taken to ensure that important lower level risks are not ignored.

Collectively, the activities of risk evaluation, categorization, and prioritization are sometimes called "risk assessment" or "risk analysis."

### Typical Work Products

1. List of risks, with a priority assigned to each risk

### Subpractices

1. Evaluate the identified risks using the defined risk parameters.

   Each risk is evaluated and assigned values in accordance with the defined risk parameters, which may include likelihood, consequence (severity, or impact), and thresholds. The assigned risk parameter values can be integrated to produce additional measures, such as risk exposure, which can be used to prioritize risks for handling.

   Often, a scale with three to five values is used to evaluate both likelihood and consequence. Likelihood, for example, can be categorized as remote, unlikely, likely, highly likely, or a near certainty.

   Examples for consequences include the following:

   - Low
   - Medium
   - High
   - Negligible
   - Marginal
   - Significant
   - Critical
   - Catastrophic

Probability values are frequently used to quantify likelihood. Consequences are generally related to cost, schedule, environmental impact, or human measures (e.g., labor hours lost and severity of injury).

This evaluation is often a difficult and time-consuming task. Specific expertise or group techniques may be needed to assess the risks and gain confidence in the prioritization. In addition, priorities may require reevaluation as time progresses.

2. Categorize and group risks according to the defined risk categories.

Risks are categorized into the defined risk categories, providing a means to look at risks according to their source, taxonomy, or project component. Related or equivalent risks may be grouped for efficient handling. The cause-and-effect relationships between related risks are documented.

3. Prioritize risks for mitigation.

A relative priority is determined for each risk based on the assigned risk parameters. Clear criteria should be used to determine the risk priority. The intent of prioritization is to determine the most effective areas to which resources for mitigation of risks can be applied with the greatest positive impact to the project.

## SG 3    Mitigate Risks

***Risks are handled and mitigated, where appropriate, to reduce adverse impacts on achieving objectives.***

The steps in handling risks include developing risk-handling options, monitoring risks, and performing risk-handling activities when defined thresholds are exceeded. Risk mitigation plans are developed and implemented for selected risks to proactively reduce the potential impact of risk occurrence. This can also include contingency plans to deal with the impact of selected risks that may occur despite attempts to mitigate them. The risk parameters used to trigger risk-handling activities are defined by the risk management strategy.

## SP 3.1    Develop Risk Mitigation Plans

***Develop a risk mitigation plan for the most important risks to the project as defined by the risk management strategy.***

A critical component of a risk mitigation plan is to develop alternative courses of action, workarounds, and fallback positions, with a recommended course of action for each critical risk. The risk mitigation plan for a given risk includes techniques and methods used to avoid, reduce, and control the probability of occurrence of the risk, the extent of damage incurred should the risk occur (sometimes called a "contingency plan"), or both. Risks are monitored and when they exceed the established thresholds, the risk mitigation plans are deployed to return the impacted effort to an acceptable risk level. If the risk cannot be mitigated, a contingency plan can be invoked. Both risk mitigation and contingency plans are often generated only for selected risks where the consequences of the risks are determined to be high or unacceptable; other risks may be accepted and simply monitored.

Options for handling risks typically include alternatives such as the following:

- Risk avoidance: Changing or lowering requirements while still meeting the user's needs

- Risk control: Taking active steps to minimize risks

- Risk transfer: Reallocating requirements to lower the risks

- Risk monitoring: Watching and periodically reevaluating the risk for changes to the assigned risk parameters

- Risk acceptance: Acknowledgment of risk but not taking any action

Often, especially for high risks, more than one approach to handling a risk should be generated.

For example, in the case of an event that disrupts continuity of operations, approaches to risk management can include the following:

- Resource reserves to respond to disruptive events

- Lists of appropriate back-up equipment to be available

- Back-up personnel for key personnel

- Plans and results of/for testing emergency response systems

- Posted procedures for emergencies

- Disseminated lists of key contacts and information resources for emergencies

In many cases, risks will be accepted or watched. Risk acceptance is usually done when the risk is judged too low for formal mitigation, or when there appears to be no viable way to reduce the risk. If a risk is accepted, the rationale for this decision should be documented. Risks are watched when there is an objectively defined, verifiable, and documented threshold of performance, time, or risk exposure (the combination of likelihood and consequence) that will trigger risk mitigation planning or invoke a contingency plan if it is needed.

Adequate consideration should be given early to technology demonstrations, models, simulations, pilots, and prototypes as part of risk mitigation planning.

**Typical Work Products**

1. Documented handling options for each identified risk

2. Risk mitigation plans

3. Contingency plans

4. List of those responsible for tracking and addressing each risk

**Subpractices**

1. Determine the levels and thresholds that define when a risk becomes unacceptable and triggers the execution of a risk mitigation plan or a contingency plan.

Risk level (derived using a risk model) is a measure combining the uncertainty of reaching an objective with the consequences of failing to reach the objective.

Risk levels and thresholds that bound planned or acceptable performance must be clearly understood and defined to provide a means with which risk can be understood. Proper categorization of risk is essential for ensuring appropriate priority based on severity and the associated management response. There may be multiple thresholds employed to initiate varying levels of management response. Typically, thresholds for the execution of risk mitigation plans are set to engage before the execution of contingency plans.

2.  Identify the person or group responsible for addressing each risk.

3.  Determine the cost-to-benefit ratio of implementing the risk mitigation plan for each risk.

Risk mitigation activities should be examined for the benefits they provide versus the resources they will expend. Just like any other design activity, alternative plans may need to be developed and the costs and benefits of each alternative assessed. The most appropriate plan is then selected for implementation. At times the risk may be significant and the benefits small, but the risk must be mitigated to reduce the probability of incurring unacceptable consequences.

4.  Develop an overall risk mitigation plan for the project to orchestrate the implementation of the individual risk mitigation and contingency plans.

The complete set of risk mitigation plans may not be affordable. A tradeoff analysis should be performed to prioritize the risk mitigation plans for implementation.

5.  Develop contingency plans for selected critical risks in the event their impacts are realized.

Risk mitigation plans are developed and implemented as needed to proactively reduce risks before they become problems. Despite best efforts, some risks may be unavoidable and will become problems that impact the project. Contingency plans can be developed for critical risks to describe the actions a project may take to deal with the occurrence of this impact. The intent is to define a proactive plan for handling the risk, either to reduce the risk (mitigation) or respond to the risk (contingency), but in either event to manage the risk.

Some risk management literature may consider contingency plans a synonym or subset of risk mitigation plans. These plans also may be addressed together as risk-handling or risk action plans.

### SP 3.2   Implement Risk Mitigation Plans

*Monitor the status of each risk periodically and implement the risk mitigation plan as appropriate.*

To effectively control and manage risks during the work effort, follow a proactive program to regularly monitor risks and the status and results of risk-handling actions. The risk management strategy defines the intervals at which the risk status should be revisited. This activity may result in the discovery of new risks or new risk-handling options that can require replanning and reassessment. In either event, the acceptability thresholds associated with the risk should be compared against the status to determine the need for implementing a risk mitigation plan.

**Typical Work Products**

1. Updated lists of risk status

2. Updated assessments of risk likelihood, consequence, and thresholds

3. Updated lists of risk-handling options

4. Updated list of actions taken to handle risks

5. Risk mitigation plans

**Subpractices**

1. Monitor risk status.

   After a risk mitigation plan is initiated, the risk is still monitored. Thresholds are assessed to check for the potential execution of a contingency plan.

   A periodic mechanism for monitoring should be employed.

2. Provide a method for tracking open risk-handling action items to closure.

   *Refer to the Project Monitoring and Control process area for more information about tracking action items.*

3. Invoke selected risk-handling options when monitored risks exceed the defined thresholds.

   Quite often, risk handling is only performed for those risks judged to be "high" and "medium." The risk-handling strategy for a given risk may include techniques and methods to avoid, reduce, and control the likelihood of the risk or the extent of damage incurred should the risk (anticipated event or situation) occur or both. In this context, risk handling includes both risk mitigation plans and contingency plans.

   Risk-handling techniques are developed to avoid, reduce, and control adverse impact to project objectives and to bring about acceptable outcomes in light of probable impacts. Actions generated to handle a risk require proper resource loading and scheduling within plans and baseline schedules. This replanning effort needs to closely consider the effects on adjacent or dependent work initiatives or activities.

   *Refer to the Project Monitoring and Control process area for more information about revising the project plan.*

4. Establish a schedule or period of performance for each risk-handling activity that includes the start date and anticipated completion date.

5. Provide continued commitment of resources for each plan to allow successful execution of the risk-handling activities.

6. Collect performance measures on the risk-handling activities.

## Generic Practices by Goal

---

**Continuous Only**

**GG 1    Achieve Specific Goals**

*The process supports and enables achievement of the specific goals of the process area by transforming identifiable input work products to produce identifiable output work products.*

**GP 1.1    Perform Specific Practices**

*Perform the specific practices of the risk management process to develop work products and provide services to achieve the specific goals of the process area.*

**GG 2    Institutionalize a Managed Process**

*The process is institutionalized as a managed process.*

---

**Staged Only**

**GG 3**      **Institutionalize a Defined Process**

*The process is institutionalized as a defined process.*

This generic goal's appearance here reflects its location in the
staged representation.

---

**GP 2.1**      **Establish an Organizational Policy**

*Establish and maintain an organizational policy for planning and
performing the risk management process.*

Elaboration:

This policy establishes organizational expectations for defining a risk
management strategy and identifying, analyzing, and mitigating risks.

**GP 2.2**      **Plan the Process**

*Establish and maintain the plan for performing the risk
management process.*

Elaboration:

This plan for performing the risk management process can be included
in (or referenced by) the project plan, which is described in the Project
Planning process area. The plan called for in this generic practice would
address the comprehensive planning for all of the specific practices in
this process area. In particular, this plan provides the overall approach
for risk mitigation, but is distinct from mitigation plans (including
contingency plans) for specific risks. In contrast, the risk mitigation plans
called for in the specific practices would address more focused items
such as the levels that trigger risk-handling activities.

**GP 2.3**      **Provide Resources**

*Provide adequate resources for performing the risk
management process, developing the work products, and
providing the services of the process.*

Elaboration:

> Examples of resources provided include the following tools:
>
> - Risk management databases
> - Risk mitigation tools
> - Prototyping tools
> - Modeling and simulation

**GP 2.4**        **Assign Responsibility**

*Assign responsibility and authority for performing the process, developing the work products, and providing the services of the risk management process.*

**GP 2.5**        **Train People**

*Train the people performing or supporting the risk management process as needed.*

Elaboration:

> Examples of training topics include the following:
>
> - Risk management concepts and activities (e.g., risk identification, evaluation, monitoring, and mitigation)
> - Measure selection for risk mitigation

**GP 2.6**        **Manage Configurations**

*Place designated work products of the risk management process under appropriate levels of control.*

Elaboration:

> Examples of work products placed under control include the following:
>
> - Risk management strategy
> - Identified risk items
> - Risk mitigation plans

**GP 2.7**        **Identify and Involve Relevant Stakeholders**

*Identify and involve the relevant stakeholders of the risk management process as planned.*

Elaboration:

> Examples of activities for stakeholder involvement include the following:
>
> - Establishing a collaborative environment for free and open discussion of risk
> - Reviewing the risk management strategy and risk mitigation plans
> - Participating in risk identification, analysis, and mitigation activities
> - Communicating and reporting risk management status

## GP 2.8     Monitor and Control the Process

***Monitor and control the risk management process against the plan for performing the process and take appropriate corrective action.***

Elaboration:

> Examples of measures and work products used in monitoring and controlling include the following:
>
> - Number of risks identified, managed, tracked, and controlled
> - Risk exposure and changes to the risk exposure for each assessed risk, and as a summary percentage of management reserve
> - Change activity for the risk mitigation plans (e.g., processes, schedule, and funding)
> - Occurrence of unanticipated risks
> - Risk categorization volatility
> - Comparison of estimated versus actual risk mitigation effort and impact
> - Schedule for risk analysis activities
> - Schedule of actions for a specific mitigation

## GP 2.9     Objectively Evaluate Adherence

***Objectively evaluate adherence of the risk management process against its process description, standards, and procedures, and address noncompliance.***

Elaboration:

> Examples of activities reviewed include the following:
>
> - Establishing and maintaining a risk management strategy
> - Identifying and analyzing risks
> - Mitigating risks

> Examples of work products reviewed include the following:
>
> - Risk management strategy
> - Risk mitigation plans

**GP 2.10     Review Status with Higher Level Management**

*Review the activities, status, and results of the risk management process with higher level management and resolve issues.*

Elaboration:

Reviews of the project risk status are held on a periodic and event-driven basis, with appropriate levels of management, to provide visibility into the potential for project risk exposure and appropriate corrective action.

Typically, these reviews include a summary of the most critical risks, key risk parameters (such as likelihood and consequence of the risks), and the status of risk mitigation efforts.

---

**Continuous Only**

**GG 3     Institutionalize a Defined Process**

*The process is institutionalized as a defined process.*

This generic goal's appearance here reflects its location in the continuous representation.

---

**GP 3.1     Establish a Defined Process**

*Establish and maintain the description of a defined risk management process.*

**GP 3.2     Collect Improvement Information**

*Collect work products, measures, measurement results, and improvement information derived from planning and performing the risk management process to support the future use and improvement of the organization's processes and process assets.*

Elaboration:

> Examples of work products, measures, measurement results, and improvement information include the following:
>
> • Risk parameters
>
> • Risk categories
>
> • Risk status reports

---

**Continuous Only**

**GG 4**  **Institutionalize a Quantitatively Managed Process**

*The process is institutionalized as a quantitatively managed process.*

  **GP 4.1**  **Establish Quantitative Objectives for the Process**

  *Establish and maintain quantitative objectives for the risk management process, which address quality and process performance, based on customer needs and business objectives.*

  **GP 4.2**  **Stabilize Subprocess Performance**

  *Stabilize the performance of one or more subprocesses to determine the ability of the risk management process to achieve the established quantitative quality and process-performance objectives.*

**GG 5**  **Institutionalize an Optimizing Process**

*The process is institutionalized as an optimizing process.*

  **GP 5.1**  **Ensure Continuous Process Improvement**

  *Ensure continuous improvement of the risk management process in fulfilling the relevant business objectives of the organization.*

  **GP 5.2**  **Correct Root Causes of Problems**

  *Identify and correct the root causes of defects and other problems in the risk management process.*

**ANEXO 3: Risk Management Process Checklist**

# RISK MANAGEMENT PROCESS CHECKLIST

## Chapter overview

## Initiation

[ ] Assemble the risk management team
[ ] Appoint the team leader
[ ] Ensure the team has a suitable breadth of skills and experience

• **Purpose**

This chapter summarizes the main steps in a simple risk management activity in the form of a process checklist.

• **Rationale**

Checklists provide an easy way of ensuring all the steps in the risk management process have been completed.

• **Method**

The process checklist here is a very simple one, and organizations that conduct projects regularly will need to tailor it to their own management processes and method of working. Refer to the preceding chapters for detailed descriptions of each step.

## Establish the context

### Objectives

[ ] Familiarize the team with the project
[ ] Assemble documentation according to the requirement
[ ] Identify the main questions and issues of concern
[ ] Review the organizational and project environment
[ ] Specify the organization's objectives

### Stakeholders

[ ] Identify the key stakeholders and their objectives
Use the stakeholder and issues summary where appropriate
[ ] Develop a communication plan if appropriate

### Criteria

[ ] Specify the criteria, linked to the project, organizational and stakeholder objectives
[ ] Develop scales for measuring the criteria, ensuring they are compatible, where relevant, with other scales used in the organization
[ ] Develop scales for measuring likelihoods that are appropriate to the project timeframe and the risk areas of interest
[ ] Develop a matrix for combining the criteria and likelihoods to derive levels of risks
Use a simple matrix for combining them if appropriate, or develop a semi-quantitative worksheet

### Key elements

[ ] Develop an analysis structure (target 20–50 key elements, items or activities)
Use the project element summary where appropriate
[ ] Number each element, describe it and list the main assumptions
Use the project element description worksheets where appropriate, or refer to a WBS Dictionary if there is one

### Risk identification

[ ] Select an appropriate process for risk identification
[ ] For each element, identify and number the risks
Include opportunities as well as risks where appropriate
[ ] Describe each risk and list the main assumptions

[ ] Assess the implications for the project
Use the risk description and response description worksheets where appropriate

## Risk analysis

[ ] Assemble data on the risks and their consequences
Most of this will be recorded on the risk and response description worksheets
[ ] Analyse the consequences of the risks in terms of the criteria
[ ] Analyse the likelihoods of the risks arising and leading to the assessed level of consequences
[ ] Summarize the analysis for each element on the assessment summary sheet
[ ] Combine the consequence and likelihood assessments to derive levels of risk
Use the assessment summary sheets

## Risk evaluation

[] Rank risks in order of decreasing level of risk
[ ] Plot the consequence and likelihood measures on the risk contour diagram if required
[ ] Draw a risk profile if appropriate
[ ] Identify Extreme or High risks for detailed risk action planning
[ ] Identify Medium risks for management and monitoring
[ ] Identify Low risks for routine management
[ ] Specify the person responsible for ensuring each risk is treated appropriately (the 'risk owner')

## Risk treatment

### Identify feasible responses

[ ] For each Extreme or High risk, and for Medium risks if resources allow, identify and number the feasible responses
Response strategies include:
• risk reduction and risk avoidance
• impact mitigation
• risk sharing
• risk retention
[ ] Describe each response and list the main assumptions
[ ] Use response description worksheets for detailed analyses
**334** Project risk management guidelines

### Select the best responses

Take into account *all* the benefits and costs, including indirect ones
Use response selection worksheets where appropriate
[ ] Select the best responses for each risk

### Develop Risk Action Plans

[ ] Develop Risk Action Plans for all Extreme and High risks
[ ] Actions (what is to be done?)
[ ] Resource requirements (what and who?)
[ ] Responsibilities (who?)
[ ] Timing (when?)
[ ] Reporting (when and to whom?)
[ ] Use risk action summary worksheets for executive reporting
[ ] Specify risk management responses for Medium risks
Use risk action summary worksheets where appropriate

## Reporting, implementation and monitoring

[ ] For major projects, produce a formal Risk Management Plan
[ ] For other projects, collate and summarize the Risk Action Plans
[ ] Implement responses and action strategies
[ ] Monitor the implementation of the Risk Action Plans
[ ] Assign responsibilities for monitoring
[ ] Specify reporting processes, frequencies and responsibilities
[ ] Undertake periodic review and evaluation